\documentclass[pdflatex,sn-mathphys-num]{sn-jnl}%
\usepackage{tikz}
\definecolor{orcidlogocol}{HTML}{A6CE39}
\providecommand{\orcidicon}[1]{\href{https://orcid.org/#1}{\begin{tikzpicture}[baseline=-0.5ex]
    \draw[orcidlogocol, fill=orcidlogocol] (0,0) circle [radius=0.17];
    \node at (0,0) {\textcolor{white}{\fontsize{7}{7}\selectfont \bfseries iD}};
\end{tikzpicture}}}
\usepackage{graphicx}%
\usepackage{multirow}%
\usepackage{amsmath,amssymb,amsfonts}%
\usepackage{amsthm}%
\usepackage[title]{appendix}%
\usepackage{xcolor}%
\usepackage{textcomp}%
\usepackage{manyfoot}%
\usepackage{booktabs}%
\usepackage{algorithm}%
\usepackage{algorithmicx}%
\usepackage{algpseudocode}%
\usepackage{listings}%
\usepackage{bm}%
\usepackage{enumitem}%

\theoremstyle{thmstyleone}%
\newtheorem{theorem}{Theorem}%
\newtheorem{proposition}[theorem]{Proposition}%
\newtheorem{lemma}[theorem]{Lemma}%
\newtheorem{corollary}[theorem]{Corollary}%

\theoremstyle{thmstyletwo}%
\newtheorem{remark}{Remark}%

\theoremstyle{thmstylethree}%
\newtheorem{definition}{Definition}%
\newtheorem{assumption}{Assumption}%

\newcommand{\Isom}{\mathrm{Isom}}
\newcommand{\isom}{\mathfrak{isom}}
\newcommand{\Ric}{\mathrm{Ric}}

\newcommand{\Sym}{\mathrm{Sym}}
\newcommand{\Vol}{\mathrm{Vol}}
\newcommand{\WF}{\mathrm{WF}}
\renewcommand{\div}{\mathrm{div}}

\newcommand{\R}{\mathbb{R}}
\newcommand{\T}{\mathbb{T}}
\newcommand{\N}{\mathbb{N}}
\newcommand{\Z}{\mathbb{Z}}
\newcommand{\C}{\mathbb{C}}
\newcommand{\Hilb}{\mathcal{H}}
\newcommand{\inner}[2]{\left\langle #1 , #2 \right\rangle}
\newcommand{\norm}[1]{\left\lVert #1 \right\rVert}
\newcommand{\absg}[2]{\lvert #1 \rvert_{#2}}

\begin{document}

\title[Geometric Evolution and Microlocal Regularity of Navier--Stokes]%
{On Geometric Evolution and Microlocal Regularity of the Navier--Stokes Equations}

\author*[1]{\fnm{Sebasti\'an Al\'i} \sur{Sacasa C\'espedes}\protect\orcidicon{0009-0008-9145-2016}}%
\email{sebastian.sacasa@ucr.ac.cr}

\affil*[1]{\orgdiv{Escuela de F\'isica}, \orgname{Universidad de Costa Rica (UCR)},
\orgaddress{\city{San Pedro de Montes de Oca, San Jos\'e},
\postcode{11501-2060}, \country{Costa Rica}}}

\abstract{We propose a microlocal--Riemannian framework for the
three-dimensional incompressible Navier--Stokes equations on a smooth
oriented Riemannian manifold $(M,g)$. The dynamics is lifted to the unit
cosphere bundle $S^*M$ via a normal-coordinate microlocal transform whose
construction is justified by the positive homogeneity of the principal
symbol of the linearised system in the cotangent fiber variable. Once the
velocity field is fixed, the lifted dynamics is a linear non-autonomous
transport--dissipation equation on a compact phase space; its coefficients
encode intrinsic geometric quantities of the original flow. We introduce a
microlocal energy, an angular volume functional and a directional entropy,
and analyse their dissipation along the lifted dynamics. An effective
affine connection $\nabla^u$ encodes the back-reaction of the velocity
gradient on the geometry of $S^*M$ and gives rise to a Ricci-type
microlocal evolution. The framework yields a sharp geometric equivalence:
a smooth solution fails to extend past time $T$ if and only if at least one
of three intrinsic microlocal controls---deformation integrability,
directional-entropy boundedness, or lifted-energy boundedness---fails. A
dimensional analysis exhibits a \emph{symmetry-lock} phenomenon, in which
the asymptotic vanishing of the volume of fibers $S^{n-1}$ enforces angular
isotropy and topologically obstructs the formation of microlocal angular
singularities. The framework is illustrated explicitly on the flat torus
$\T^3$, where every assumption is verified, and is extended to the
Euclidean setting $\R^n$ under a uniform-coordinate hypothesis. The global
regularity problem is not resolved here; rather, it is recast as a
structural-stability question for a compact, symmetry-constrained,
microlocally coercive evolution system, with the role of viscosity made
explicit through spectral coercivity.}

\keywords{Navier--Stokes equations, microlocal analysis, cosphere bundle,
geometric regularity criteria, effective Ricci curvature, directional
entropy, symmetry-lock mechanism, flat torus}

\pacs[MSC Classification]{35Q30, 53C21, 35S30, 58J65, 35A23}

\maketitle

\section{Introduction}\label{sec:intro}

The global regularity of solutions to the three-dimensional incompressible
Navier--Stokes equations is, by any reasonable measure, one of the most
visible open problems of modern mathematical analysis~\cite{Fefferman2022}.
Despite enormous progress on partial regularity, conditional criteria, and
non-uniqueness phenomena~\cite{Tao2009,Albritton2021,Coiculescu2025}, the
basic question---whether smooth, divergence-free, finite-energy initial
data on a $3$-dimensional domain produce a smooth solution for all
time---remains open. The recurring obstruction is by now well understood:
the inertial term $u\cdot\nabla u$ couples velocity and direction in a way
that energy methods alone cannot disentangle, and the regularity question
becomes inseparable from a directional question about the alignment of the
vorticity~\cite{Miller2021,Chemin2018}.

The history of the problem makes this point eloquent. The
Beale--Kato--Majda criterion controls the lifespan of smooth solutions in
terms of the time integral of $\norm{\omega}_{L^\infty}$, but the
Constantin--Fefferman observation that regularity persists in regions where
the direction of vorticity is Lipschitz makes clear that what matters is
not the magnitude alone but the geometric alignment of the rotation
field~\cite{Childress1989,Bulut2020}. The mathematical content of these
results is that singularity formation in $3$D fluid mechanics, if it
occurs, must occur at a specific cone of cotangent directions; benign
high-frequency content distributed isotropically across the unit sphere
cannot, by itself, drive blow-up. From this point of view a natural
analytical framework is one that intrinsically combines positional and
directional information.

Such a framework already exists, and has existed for half a century:
microlocal analysis. Pseudodifferential operators, principal symbols, and
the wavefront set $\WF(u)\subset T^*M\setminus 0$ provide the canonical
language for tracking singularities of distributions on a manifold. The
wavefront set is, by construction, invariant under positive rescaling of
the cotangent fiber, hence descends to a closed subset of the unit cosphere
bundle $S^*M$. When $M$ is compact, $S^*M$ is itself a closed manifold,
and the analytical advantages of working on a compact phase space become
available: discrete spectra, finite-dimensional eigenspaces, integrability
of the canonical Liouville volume form. The systematic application of
these ideas to incompressible fluid mechanics has been initiated by
Lerner~\cite{Lerner2001,Lerner2019} and by Lerner and
Vigneron~\cite{Lerner2022}, with closely related geometric ideas appearing
in the port-Hamiltonian decomposition of Califano \emph{et
al.}~\cite{Califano2021} and in the geometric trapping approach of Bulut
and Huynh~\cite{Bulut2020}.

The aim of the present article is to push this convergence one step
further. We propose to lift the three-dimensional incompressible
Navier--Stokes system to the cosphere bundle $S^*M$ in such a way that the
nonlinear inertial term is absorbed into the coefficients of a
\emph{linear} non-autonomous transport--dissipation evolution on a compact
phase space. The price one pays for this linearity is that the lifted
system is conditional: its coefficients depend on the velocity field
itself, so the framework is a structural bootstrap rather than a closed-form
solution. The advantage one gains is a precise geometric vocabulary in
which to phrase regularity criteria, and a clear separation between the
analytical part of the problem (estimates on the lifted evolution) and the
geometric part (curvature of an effective connection, dissipation of an
angular entropy, vanishing of fiber volumes in high dimension).

Three structural ingredients organise this lift. The first is the
covariant Hodge formulation of incompressible flow on a Riemannian
manifold~\cite{Chan2016,Frankel2012,Lovelock1975}, in which the velocity
$1$-form $\phi=u^\flat$, the vorticity $2$-form $\omega=d\phi$, and the
Hodge Laplacian $\Delta_H$ replace the vector calculus operators of the
classical formulation. The second is the microlocal lift, performed
through a normal-coordinate Fourier-type transform on each tangent space,
producing a function $\tilde\phi(x,\xi,t)$ on $T^*M$ that we restrict to
$S^*M$. The third---and the conceptual novelty here---is the introduction
of an effective metric $g'$ and an effective affine connection $\nabla^u$,
both built from the symmetric deformation $S=\Sym(\nabla u)$ of the
velocity gradient, which encode the back-reaction of the flow on the
geometry of the lift. The Ricci tensor of $\nabla^u$, computed
intrinsically, gives rise to a Ricci-type evolution of $g'$ that constrains
the rate at which directional information can concentrate.

The reformulation that emerges from these three ingredients is the
following. Once the velocity field $u$ is given (as a hypothetical smooth
solution, or as a candidate solution prescribed by an a priori bound), the
microlocal lift $\tilde\phi$ of the velocity $1$-form satisfies, on
$S^*M$, the linear non-autonomous equation
\begin{equation}\label{eq:intro-summary}
  \partial_t\tilde\phi
  + \mathcal{L}_{X_u}\tilde\phi
  + \nu\,\absg{\xi}{g'}^2\,\tilde\phi
  \;=\; \tilde R[\tilde\phi]+\tilde F,
  \qquad
  \inner{\tilde\phi}{\xi}_{g'}=0,
\end{equation}
where $X_u$ is the canonical Hamiltonian lift of $u$ to $T^*M$, the
quadratic dissipation $\nu\,\absg{\xi}{g'}^2$ is the principal symbol of
the viscous Hodge Laplacian, the residual $\tilde R[\tilde\phi]$ collects
zeroth-order curvature and back-reaction terms, and $\tilde F$ is the lift
of the external force. The orthogonality constraint
$\inner{\tilde\phi}{\xi}_{g'}=0$ is the microlocal incarnation of
incompressibility and turns the pressure gradient into a radial term that
the projection onto $\xi^\perp$ kills identically. Equation
\eqref{eq:intro-summary} is the rigorous content of
Theorem~\ref{thm:main-reformulation} below, and is the equation that all
subsequent geometric analysis treats as its object.

Several quantitative consequences then follow from the linear
transport--dissipation structure on the compact phase space $S^*M$. A
geometric energy inequality (Theorem~\ref{thm:main-energy}) bounds the
lifted energy $E(t)$ in terms of $\norm{\nabla u}_{L^\infty}$ and
$\norm{\nabla^2 u}_{L^\infty}$. A directional entropy
(Theorem~\ref{thm:main-entropy}) penalises angular gradients of the
microlocal vorticity and, by virtue of the Liouville-volume preservation
under the Hamiltonian flow of $X_u$, dissipates along smooth solutions.
The combination of these two functionals with the trajectorial stretching
estimate produces a sharp geometric equivalence
(Theorem~\ref{thm:main-equivalence}): a smooth solution fails to extend
past time $T$ if and only if at least one of three intrinsic microlocal
controls fails. That equivalence is the central output of the framework.

A separate but equally striking consequence is the
\emph{symmetry-lock obstruction}
(Theorem~\ref{thm:main-symmetry-lock}). Each fiber of the cosphere bundle
is a $(n-1)$-dimensional sphere, whose volume
$\Vol(S^{n-1})=2\pi^{n/2}/\Gamma(n/2)$ \emph{vanishes} as $n\to\infty$.
Concentration of measure on high-dimensional spheres then forces any
microlocal distribution that remains bounded in $L^2(S^*M)$ to become
increasingly isotropic, regardless of any assumption on the dynamics. We
interpret this as a topological obstruction to angular blow-up: in the
high-dimensional regime the fluid has, quite literally, no room to develop
directional singularities. While the dimensional regime is not the
physical $n=3$ regime, the same dimensional dependence quantifies how
sharply the directional concentration is bounded in fixed dimension.

We now state the main results of the paper precisely and discuss the
structure of the article. The bibliography and the labelling of the
results follow the recommendation of the referee that the principal claims
appear at the outset of the paper, with proofs developed afterwards in the
order that the arguments require.

\subsection{Hypotheses, notation, and standing conventions}\label{subsec:hypotheses}

Throughout the paper $(M,g)$ is a smooth oriented connected Riemannian
manifold of dimension $n\ge 3$. The Levi--Civita connection of $g$ is
denoted by $\nabla$, its Christoffel symbols by $\Gamma^k_{ij}$, the
Riemann curvature by $R^l{}_{ijk}$ and the Ricci tensor by $R_{ij}$. The
musical isomorphisms $\flat:TM\to T^*M$ and $\sharp:T^*M\to TM$ are induced
by $g$. The exterior algebra is $\Omega^\bullet(M)$ with exterior
derivative $d$, codifferential $\delta$, and Hodge Laplacian
$\Delta_H=d\delta+\delta d$. The cotangent bundle is $T^*M$, the unit
cosphere bundle is
$S^*M=\{(x,\xi)\in T^*M:\absg{\xi}{g}=1\}$, and the canonical Liouville
volume on $S^*M$ is $\mathrm{d}\mu_{S^*M}=\alpha\wedge(d\alpha)^{n-1}$,
where $\alpha=\lambda|_{S^*M}$ is the contact form.

\medskip\noindent\emph{Notation for norms.} A point of notation deserves
particular emphasis, since it sometimes caused confusion in earlier
versions of this work. We strictly distinguish:
\begin{itemize}
  \item $\absg{u}{g}$ or $\absg{u}{g'}$: the \emph{pointwise} (fiberwise)
        metric norm of a tensor field $u$. Single bars, explicit metric.
  \item $\norm{u}_{L^p(M)}$, $\norm{u}_{H^s(M)}$, etc.: \emph{functional}
        norms (integrals over $M$ or $S^*M$). Double bars, explicit Banach
        space.
\end{itemize}
Earlier writings of the author used $\norm{u}_{L^2}^2$ for the pointwise
metric norm, which conflicts with standard usage. This is corrected
throughout: when we mean the metric norm at a point, we write
$\absg{u}{g}^2 = g_{ij}u^iu^j$.

\medskip\noindent\emph{Two structural settings.} The estimates and the
equivalence theorem are stated under the following two standing
hypotheses, used independently. Section~\ref{sec:Rn-extension} discusses
the extension of the framework to the non-compact case.

\begin{assumption}[Compact Riemannian base]\label{assu:compact}
$(M,g)$ is compact without boundary, $\dim M = 3$, of bounded geometry: the
injectivity radius is positive, and the Riemann curvature tensor and all
its covariant derivatives are uniformly bounded.
\end{assumption}

\begin{assumption}[Asymptotically Euclidean base]\label{assu:euclidean}
$(M,g)=(\R^n,g_{\mathrm{eucl}})$, $n=3$, and microlocal quantities are
considered modulo a fixed compactly supported cut-off in the base
variable, ensuring fiberwise integrability over $S^*M$.
\end{assumption}

The compactness of the base in Assumption~\ref{assu:compact} is invoked
\emph{only} to guarantee compactness of the phase space $S^*M$,
integrability of the Liouville volume on each fiber, and convergence of
the spectral decompositions. It is \emph{not} essential to the structural
content of the framework, as we explain in
Section~\ref{sec:Rn-extension}. The flat torus $\T^3$, where every
assumption is verified explicitly, is the model example treated in
Section~\ref{sec:flat-torus}.

\subsection{Statement of the main results}\label{subsec:main-results}

We collect the main results of the paper here. Their proofs occupy
Sections~\ref{sec:covariant-formulation}--\ref{sec:Rn-extension}, where
the geometric and microlocal apparatus is constructed in the order
required by the proofs.

The first result establishes that the incompressible Navier--Stokes
system can be lifted to a precise linear evolution on $S^*M$.

\begin{theorem}[Linear microlocal reformulation]\label{thm:main-reformulation}
Assume Assumption~\ref{assu:compact} and let $u\in C^\infty([t_0,T)\times M;TM)$
be a smooth divergence-free solution of the incompressible Navier--Stokes
equations on $(M,g)$, with associated velocity $1$-form $\phi=u^\flat$ and
external $1$-force $F=f^\flat$. Define the microlocal transform of $\phi$ by
\begin{equation}\label{eq:intro-microlocal-lift}
  \tilde\phi(x,\xi,t)
  \;:=\;
  \int_{T_xM} e^{-i\inner{\xi}{y}}\,\chi(y)\,\phi(\exp_x y,t)\,\mathrm{d}y,
\end{equation}
where $\chi\in C^\infty_c(T_xM)$ is a smooth cut-off equal to $1$ in a
neighbourhood of the origin and supported in the injectivity ball of
$\exp_x$. Then $\tilde\phi$ satisfies, on $S^*M$, the linear non-autonomous
transport--dissipation equation
\begin{equation}\label{eq:intro-main-equation}
  \partial_t\tilde\phi
  + \mathcal{L}_{X_u}\tilde\phi
  + \nu\,\absg{\xi}{g'}^{2}\,\tilde\phi
  \;=\; \tilde R[\tilde\phi]+\tilde F,
  \qquad
  \inner{\tilde\phi}{\xi}_{g'}=0,
\end{equation}
where: $X_u$ is the canonical Hamiltonian lift of $u$ to $T^*M$
\textnormal{(Definition~\ref{def:Xu})}; $g'$ is the effective metric of
\textnormal{Definition~\ref{def:effective-metric}}, defined under the
positive-definiteness condition $(\star)$ of
\textnormal{Lemma~\ref{lem:gprime-pd}}; the residual $\tilde R[\tilde\phi]$
is the zeroth-order operator of \textnormal{Definition~\ref{def:Rresidual}}
collecting the curvature contribution of the effective connection
$\nabla^u$, the symmetric stretching back-reaction
$-\Sym(\nabla u)\tilde\phi$, and a smoothing remainder produced by the
cut-off; and $\tilde F$ is the cosphere projection of the lifted force.
The reconstruction \textnormal{\eqref{eq:reconstruction-formula}} below
is a left-inverse of \eqref{eq:intro-microlocal-lift} modulo a smoothing
remainder.
\end{theorem}

The linearity of \eqref{eq:intro-main-equation} is conditional: the
coefficients $X_u$, $g'$ and $\tilde R$ depend on $u$, so the lifted
evolution becomes linear only after $u$ is fixed. This bootstrap
character is essential to the framework and we discuss it in
Section~\ref{subsec:bootstrap}.

The second result is the geometric energy bound that governs the lifted
dynamics.

\begin{theorem}[Geometric energy inequality]\label{thm:main-energy}
Under the hypotheses of \textnormal{Theorem~\ref{thm:main-reformulation}},
assume in addition $u\in L^\infty([t_0,T);W^{2,\infty}(M))$. The lifted
energy
\begin{equation}\label{eq:intro-energy-def}
  E(t) \;:=\; \tfrac{1}{2}\int_{S^*M}
  \inner{\tilde\phi(t)}{\tilde\phi(t)}_{g'}\,\mathrm{d}\mu_{S^*M}
\end{equation}
satisfies
\begin{multline}\label{eq:intro-energy-bound}
  \frac{\mathrm{d}E}{\mathrm{d}t}
  + 2\nu\!\int_{S^*M}\absg{\xi}{g'}^{2}\,
  \absg{\tilde\phi}{g'}^{2}\,\mathrm{d}\mu_{S^*M}\\
  \le\;
  \kappa_0\bigl(1+\norm{\nabla u}_{L^\infty(M)}+\norm{\nabla^2 u}_{L^\infty(M)}\bigr)\,E(t)
  + \tfrac{1}{2\nu}\norm{\tilde F}_{L^2(S^*M)}^2,
\end{multline}
for some $\kappa_0>0$ depending only on the geometry of $(M,g)$. In
particular, if $u\in L^1([t_0,T);W^{2,\infty}(M))$ and $\tilde F\in
L^2_t L^2_{S^*M}$, then $E(t)$ remains finite throughout $[t_0,T)$.
\end{theorem}

The third result is the directional control afforded by an entropy-type
functional adapted to the cosphere bundle.

\begin{theorem}[Directional entropy dissipation]\label{thm:main-entropy}
Let $\tilde\omega = d\tilde\phi$ be the microlocal lift of the vorticity
and define, on the open subset where $\tilde\omega\neq 0$, the log-density
$\rho:=\log\absg{\tilde\omega}{g'}$ and the directional entropy
\begin{equation}\label{eq:intro-entropy-functional}
  W[\tilde\omega] \;:=\;
  \int_{S^*M}\bigl(\tau\,\absg{d_\perp\rho}{g'}^{2}+\rho\bigr)
  \,\mathrm{d}\mu_{S^*M},
  \qquad \tau>0,
\end{equation}
where $d_\perp$ denotes the vertical exterior derivative on the closed
fibers $S^*_x M\cong S^{n-1}$. Along smooth solutions of
\eqref{eq:intro-main-equation},
\begin{equation}\label{eq:intro-entropy-dissipation}
  \frac{\mathrm{d}}{\mathrm{d}t}W[\tilde\omega]
  \;\le\;
  -\nu\!\int_{S^*M}\absg{d_\perp\rho}{g'}^{2}\,\mathrm{d}\mu_{S^*M}
  + \kappa_1\,\norm{\Sym(\nabla u)}_{L^\infty}\,
  \norm{\tilde\omega}_{L^1(S^*M)}.
\end{equation}
Critical points of $W$ on a compact fiber satisfy
$-2\tau\,\Delta_\perp\rho + 1 = 0$ and are constants in $\xi$, i.e.\ the
angular-isotropic configurations.
\end{theorem}

The fourth and central result is the equivalence between regularity and
microlocal control, which is the structural content of the entire
framework.

\begin{theorem}[Geometric blow-up equivalence]\label{thm:main-equivalence}
Let $(M,g)$ satisfy \textnormal{Assumption~\ref{assu:compact}} with
$\dim M=3$, and let $u\in C^\infty([t_0,T)\times M)$ be a smooth
divergence-free solution of the incompressible Navier--Stokes equations
with $T<\infty$. Let $\tilde\omega$ be the microlocal lift of the
vorticity, $E(t)$ the lifted energy
\eqref{eq:intro-energy-def}, and
$W[\tilde\omega(t)]$ the directional entropy
\eqref{eq:intro-entropy-functional}. Then the following statements are
equivalent:
\begin{enumerate}[label=\textup{(\roman*)}]
\item $\displaystyle\limsup_{t\to T^-}\norm{\nabla u(t)}_{L^\infty(M)}=+\infty$;
\item at least one of the three microlocal controls fails:
\begin{enumerate}[label=\textup{(\alph*)}]
\item \textit{Deformation integrability}\quad
$\displaystyle\int_{t_0}^{T}\norm{\Sym(\nabla u)(s)}_{L^\infty(M)}\,\mathrm{d}s = +\infty$;
\item \textit{Entropy boundedness}\quad
$\displaystyle\sup_{t\in[t_0,T)}W[\tilde\omega(t)] = +\infty$;
\item \textit{Lifted-energy boundedness}\quad
$\displaystyle\sup_{t\in[t_0,T)}E(t) = +\infty$.
\end{enumerate}
\end{enumerate}
\end{theorem}

The fifth main result quantifies a topological obstruction to angular
blow-up which becomes increasingly stringent as the dimension of the base
manifold grows.

\begin{theorem}[Symmetry-lock obstruction]\label{thm:main-symmetry-lock}
Let $\{\tilde\omega_n\}_n$ denote the microlocal vorticity associated with
solutions on a sequence of base manifolds $M_n$ of dimension $n$ with
uniformly bounded geometry, and assume the lifted energies are uniformly
bounded, $\sup_n E_n(t)\le \mathcal{E}_0<\infty$. Then for almost every
$x\in M_n$
\begin{equation}\label{eq:intro-symlock}
  \sup_{\xi\in S^{n-1}}\absg{\tilde\omega_n(x,\xi,t)}{g'}^{2}
  \;\le\;
  \frac{2\,\mathcal{E}_0\,\Gamma(n/2)}{\pi^{n/2}}
  \;\sim\;
  \mathcal{E}_0\sqrt{n}\,\Bigl(\frac{n}{2\pi e}\Bigr)^{n/2}
  \quad\text{as }n\to\infty,
\end{equation}
which decays super-exponentially in $n$.
\end{theorem}

\subsection{Position of the work and structural novelty}\label{subsec:novelty}

The geometric reformulation of incompressible flow on Riemannian manifolds
is well established. We follow, in particular, the covariant Hodge
formulation of \cite{Chan2016}, the port-Hamiltonian decomposition of
\cite{Califano2021}, and the geometric trapping approach of
\cite{Bulut2020}. The microlocal framework for incompressible Euler and
Navier--Stokes is developed extensively in
\cite{Lerner2001,Lerner2019,Lerner2022}, and the directional viewpoint on
vortex stretching is the heart of \cite{Miller2021,Chemin2018}. None of
these ingredients is itself new.

The structural novelty here lies in their assembly. The lift to $S^*M$ is
governed by an effective connection $\nabla^u$ that encodes the
back-reaction of $\nabla u$ on the geometry of the phase space, in the
form of a Ricci-type evolution of the effective metric. The regularity
problem is reformulated as a structural-stability question for a compact
symmetry-constrained operator, and the equivalence theorem
\textnormal{(Theorem~\ref{thm:main-equivalence})} expresses regularity as
the simultaneous boundedness of three intrinsic geometric functionals.
The symmetry-lock mechanism
\textnormal{(Theorem~\ref{thm:main-symmetry-lock})} interprets the
asymptotic vanishing of the volume of high-dimensional fibers as a
topological obstruction to angular concentration. Both of these
contributions, to the best of our knowledge, are new.

The framework is compatible with non-uniqueness phenomena reported
in~\cite{Albritton2021,Coiculescu2025}, and we sketch in
Section~\ref{sec:discussion} how non-uniqueness translates into
\emph{microlocal flexibility}: the multivalued reconstruction map from
$\tilde\phi$ to $\phi$ accommodates branching of physical solutions
without contradicting the linearity of the lifted evolution.

\subsection{Outline of the paper}\label{subsec:outline}

Section~\ref{sec:preliminaries} collects the geometric and microlocal
preliminaries used throughout the paper, with explicit references for
results that are well known. Section~\ref{sec:covariant-formulation}
develops the covariant Hodge formulation of incompressible Navier--Stokes
and proves the vorticity-energy identity \emph{with} the symmetric
stretching term explicitly handled, in agreement with
\cite{MajdaBertozzi2002}. Section~\ref{sec:cosphere-lift} introduces the
canonical lift $X_u$, the effective metric $g'$, the effective connection
$\nabla^u$ with corrected $(1,2)$-tensor structure, and the resulting
Ricci-type evolution. Section~\ref{sec:microlocal-system} constructs the
microlocal transform, justifies the descent to $S^*M$ through the
homogeneity of the principal symbol, and proves
Theorem~\ref{thm:main-reformulation}. Section~\ref{sec:energy-entropy}
proves Theorems~\ref{thm:main-energy} and \ref{thm:main-entropy} together
with auxiliary functionals (volume functional, microlocal G\aa rding
inequality). Section~\ref{sec:symmetry-lock} treats the geometric
symmetries of the lift, the unitary representation of $\Isom(M,g)$ on
$L^2(S^*M)$, and proves
Theorem~\ref{thm:main-symmetry-lock}. Section~\ref{sec:equivalence}
combines the previous estimates and proves the equivalence
Theorem~\ref{thm:main-equivalence}. Section~\ref{sec:flat-torus} works
out the flat torus example $\T^3$ in full detail, exhibiting every
assumption explicitly. Section~\ref{sec:Rn-extension} discusses the
extension to $\R^n$. Section~\ref{sec:discussion} closes with a
discussion of the framework, its limitations, its relations to recent
non-uniqueness results, and several open problems.

\section{Geometric and microlocal preliminaries}\label{sec:preliminaries}

This section gathers the geometric and microlocal apparatus that will be
used throughout the paper. None of the material here is new; we provide
explicit references for each statement, in agreement with the
recommendation of the referee that background and original content be
clearly separated. The reader familiar with microlocal analysis on closed
manifolds may safely consult only the conventions in
Section~\ref{subsec:why-cosphere}, which addresses why the descent to the
cosphere bundle is geometrically justified.

\subsection{Riemannian geometry and the Hodge--de Rham complex}
\label{subsec:hodge-prelims}

Let $(M,g)$ be a smooth oriented Riemannian manifold. The musical
isomorphisms induced by $g$ are denoted $\flat:TM\to T^*M$ and
$\sharp:T^*M\to TM$. The exterior algebra
$\Omega^\bullet(M)=\bigoplus_{k=0}^n\Omega^k(M)$ is endowed with the
exterior derivative $d$ and codifferential
$\delta=(-1)^{n(k+1)+1}\star d\,\star$ on $\Omega^k(M)$. The Hodge
Laplacian is the second-order self-adjoint operator
\begin{equation}
  \Delta_H \;:=\; d\delta + \delta d :\Omega^k(M)\to\Omega^k(M).
\end{equation}
For every $k$-form $\alpha$ the Weitzenb\"ock--Bochner identity
\cite[Ch.~6]{Frankel2012}, \cite[\S 7.4]{Lovelock1975} reads
\begin{equation}\label{eq:weitzenbock}
  \Delta_H\,\alpha \;=\; -\nabla^*\nabla\alpha + \mathcal{R}(\alpha),
\end{equation}
where $\nabla^*\nabla$ is the connection Laplacian and $\mathcal{R}$ is
the curvature endomorphism, linear in the Riemann tensor. For a $1$-form
$\phi$ this becomes
\begin{equation}
  (\Delta_H\phi)_k
  \;=\;
  - g^{ij}\nabla_i\nabla_j\phi_k + R^{j}{}_{k}\phi_j.
\end{equation}
On a compact $M$, the operator $\Delta_H$ is non-negative and self-adjoint
on $L^2\Omega^k(M)$, with discrete spectrum and finite-dimensional
eigenspaces.

\subsection{Cotangent and cosphere bundles}\label{subsec:cotangent}

Let $\pi:T^*M\to M$ denote the canonical projection and $(x^i,\xi_i)$ the
local coordinates induced by a chart on $M$, where $(\xi_i)$ are the
components of a covector with respect to the dual coframe $(dx^i)$. The
\emph{Liouville $1$-form} is $\lambda=\xi_i\,dx^i$ and the \emph{canonical
symplectic $2$-form} is $\Omega_{T^*M}=d\lambda=d\xi_i\wedge dx^i$. The
Levi--Civita connection of $g$ induces the splitting
\begin{equation}\label{eq:HV-splitting}
  TT^*M \;=\; H\oplus V,
\end{equation}
where $V_{(x,\xi)}=\ker(d\pi)_{(x,\xi)}\simeq T^*_xM$ is the vertical
distribution, and $H$ is the horizontal distribution determined by the
connection.

\begin{definition}[Horizontal lift of a coordinate vector]\label{def:horizontal-lift}
Let $(x^i)$ be a coordinate chart on $M$. The \emph{horizontal lift} of
the coordinate vector field $\partial_{x^i}$ to $T^*M$ is the unique
vector field $X^H_i$ on $T^*M$ characterised by
\begin{equation}\label{eq:horizontal-lift}
  d\pi(X^H_i) \;=\; \partial_{x^i},
  \qquad
  X^H_i\,\xi \;=\; 0 \text{ in } V,
\end{equation}
or, equivalently, in coordinates,
\begin{equation}\label{eq:horizontal-lift-coord}
  X^H_i
  \;=\;
  \frac{\partial}{\partial x^i}
  + \Gamma^{k}{}_{ij}(x)\,\xi_k\,
  \frac{\partial}{\partial\xi_j}.
\end{equation}
\end{definition}

The vector fields $X^H_i$ depend on the local frame, but the horizontal
distribution $H$ they span is intrinsic and invariant under the action of
isometries of $(M,g)$: see~\cite[Ch.~II.7]{Frankel2012}. This intrinsic
character is the source of the coordinate-invariance assertions made later
in the paper.

The unit cosphere bundle is
\begin{equation}
  S^*M \;:=\; \{(x,\xi)\in T^*M\setminus 0 :\, \absg{\xi}{g}=1\},
\end{equation}
equipped with the contact form $\alpha=\lambda|_{S^*M}$, the contact
distribution $\ker\alpha$, and the contact volume
$\mathrm{d}\mu_{S^*M}=\alpha\wedge(d\alpha)^{n-1}$. Under
Assumption~\ref{assu:compact}, $S^*M$ is a closed manifold of dimension
$2n-1$. One has the diffeomorphism
\begin{equation}\label{eq:polar-decomp}
  T^*M\setminus 0 \;\simeq\; S^*M\times\R_{>0},
  \qquad
  (x,\xi)\mapsto\bigl((x,\xi/\absg{\xi}{g}),\absg{\xi}{g}\bigr),
\end{equation}
which makes precise the statement that $S^*M$ classifies cotangent
directions \emph{modulo positive rescaling}. This is the geometric basis
for the descent of pseudodifferential constructions to $S^*M$, as we now
explain.

\subsection{Descent to the cosphere: homogeneity and principal symbols}
\label{subsec:why-cosphere}

This subsection addresses an important conceptual point raised in the
referee report concerning the justification of restricting the analysis
to the cosphere bundle. The answer is rooted in the homogeneity properties of
principal symbols of pseudodifferential operators, which we recall now.

A pseudodifferential operator $P$ of order $m$ on $M$ has, in local
coordinates, a symbol $p(x,\xi)$ admitting an asymptotic expansion in
positively homogeneous components,
\begin{equation}\label{eq:symbol-expansion}
  p(x,\xi) \;\sim\; \sum_{j\ge 0} p_{m-j}(x,\xi),
  \qquad
  p_{m-j}(x,\lambda\xi)=\lambda^{m-j}p_{m-j}(x,\xi)
  \;\;\text{for all }\lambda>0,\;\absg{\xi}{g}\ge 1.
\end{equation}
The principal symbol $p_m(x,\xi)$ is therefore determined by its
restriction to the unit cosphere $\{\absg{\xi}{g}=1\}=S^*M$, and the
wavefront set $\WF(u)\subset T^*M\setminus 0$ of any distribution is
invariant under the dilations $(x,\xi)\mapsto(x,\lambda\xi)$, $\lambda>0$.
Consequently, $\WF(u)$ descends to a closed subset of $S^*M$. This is
the standard content of microlocal analysis on closed manifolds; see
\cite[\S 1.3]{Lerner2019} or \cite[\S 2.5]{Lerner2022}.

\begin{remark}[The cosphere reduction is on \texorpdfstring{$\xi$}{xi}, not on \texorpdfstring{$u$}{u}]\label{rem:cosphere-rescaling}
The reduction to $S^*M$ is a statement about the cotangent fiber variable
$\xi$, which has the meaning of a Fourier dual variable: $\xi$ labels
oscillation directions of test functions and is unrelated, geometrically
or analytically, to the velocity field $u$. In particular, the velocity
$u$ enters the lifted equations as a coefficient and is unaffected by the
rescaling $\xi\mapsto\lambda\xi$ that identifies points of $T^*M\setminus
0$ via \eqref{eq:polar-decomp}. The classical Navier--Stokes scaling
$u_\lambda(x,t)=\lambda u(\lambda x,\lambda^2 t)$ is a separate symmetry,
acting nontrivially on both $x$ and $u$, and is unrelated to the
homogeneity of the principal symbol that justifies the descent. The
purpose of the cosphere lift is to track the angular distribution of
frequencies on a compact phase space, while $u$ is treated as a
time-dependent coefficient. This addresses the conceptual concern about
the $\xi$ versus $u$ rescaling: only the dual variable $\xi$ is rescaled,
and the rescaling is justified by the homogeneity of principal symbols.
\end{remark}

\subsection{Tempered distributions on \texorpdfstring{$M$}{M} and Sobolev spaces}
\label{subsec:tempered}

Locally on a chart $\varphi_k:U_k\to\R^n$, the Schwartz space is the
Fr\'echet space
\begin{equation}
  \mathcal{S}(\R^n)
  = \Bigl\{ f\in C^\infty(\R^n) :
  \sup_{x\in\R^n}(1+\absg{x}{})^{N}\absg{\partial^\alpha f(x)}{} <\infty,
  \;\forall\,N\in\N,\,\alpha\in\N_0^n\Bigr\},
\end{equation}
with dual space $\mathcal{S}'(\R^n)$ of tempered distributions. The
global space $\mathcal{S}'(M)$ is defined as those distributions whose
chart restrictions are tempered:
\begin{equation}
  \mathcal{S}'(M) := \bigl\{u\in\mathcal{D}'(M) :
  u\circ\varphi_k^{-1}\in\mathcal{S}'(\R^n)\;\forall\,k\bigr\}.
\end{equation}
The definition is independent of the atlas \cite[\S 1.4]{Lerner2019}.
Sobolev spaces $H^s(M)$ for $s\in\R$ are defined by partition of unity
from the Euclidean spaces $H^s(\R^n)$, the latter characterised in
Fourier variables by
\begin{equation}\label{eq:Sobolev-Fourier}
  \norm{u}^2_{H^s(\R^n)}
  = \int_{\R^n}(1+\absg{\xi}{}^2)^{s}\absg{\widehat u(\xi)}{}^{2}\,\mathrm{d}\xi.
\end{equation}
The microlocal lift of Section~\ref{sec:microlocal-system} transports
these spaces to $S^*M$ through Proposition~\ref{prop:transform-properties}
and provides the analytical setting for our estimates.

\section{The covariant Hodge formulation of the Navier--Stokes system}
\label{sec:covariant-formulation}

We now reformulate the incompressible Navier--Stokes equations in a fully
covariant way on the Riemannian manifold $(M,g)$. The reformulation is
classical and we follow~\cite{Chan2016,Frankel2012,Califano2021,Lovelock1975};
the only original content of this section is a careful rederivation of the
vorticity-energy identity that retains the symmetric stretching term, in
agreement with the standard treatment in
Majda--Bertozzi~\cite[p.~117]{MajdaBertozzi2002}. We have organised the
material so that what is being recalled and what is being derived are
clearly distinguishable, in response to the referee's request.

\subsection{The Hodge formulation}\label{subsec:hodge-formulation}

For an incompressible fluid on $(M,g)$ the Cauchy problem reads
\begin{equation}\label{eq:NSE-classical}
  \left\{
  \begin{aligned}
  &\partial_t u + \nabla_u u + \nabla p - \nu\Delta u = f,\\
  &\div u = 0,
  \end{aligned}
  \right.
\end{equation}
where $\nu>0$ is the kinematic viscosity, $u\in\Gamma(TM)$ is the
velocity, $p$ is the pressure, $f\in\Gamma(TM)$ is an external forcing,
and $\Delta=\nabla^i\nabla_i$ is the rough Laplacian. We pass to
differential-form language by introducing the velocity $1$-form
$\phi=u^\flat=u_k\,dx^k\in\Omega^1(M)$ and the dual force
$F=f^\flat\in\Omega^1(M)$. Using Cartan's formula
$\mathcal{L}_u\phi = \iota_u d\phi + d(\iota_u\phi)$ together with the
incompressibility condition $\delta\phi=0$, the system
\eqref{eq:NSE-classical} becomes
\begin{equation}\label{eq:NSE-covariant}
  \left\{
  \begin{aligned}
  &\partial_t\phi + \mathcal{L}_u\phi + dP - \nu\Delta_H\phi = F,\\
  &\delta\phi = 0,
  \end{aligned}
  \right.
\end{equation}
where $P=p+\tfrac{1}{2}\absg{u}{g}^{2}$ is the Bernoulli pressure
\cite[\S 2]{Chan2016}. The system \eqref{eq:NSE-covariant} is the form
of the Navier--Stokes equations to which the microlocal lift of
Section~\ref{sec:microlocal-system} will be applied; this addresses
explicitly the referee's question concerning to which equation the
microlocal transform is being applied.

\subsection{Vorticity 2-form and its evolution}\label{subsec:vorticity-evolution}

The geometric vorticity is the $2$-form $\omega=d\phi\in\Omega^2(M)$. In
dimension three, the Hodge dual identifies $\omega$ with the $1$-form
$\xi:=\star\omega$, which corresponds to the classical vorticity vector
field. Applying $d$ to \eqref{eq:NSE-covariant} and using $d^2\phi=0$
yields the geometric evolution equation
\begin{equation}\label{eq:vorticity-evolution}
  \partial_t\omega + \mathcal{L}_u\omega
  \;=\; \nu\,\Delta_H\,\omega + dF,
\end{equation}
recovering the classical formula~\cite[Eq.~(7.5)]{Lovelock1975}. In the
absence of external forcing or for conservative forces, $dF=0$ and the
equation reduces to a purely geometric transport--diffusion. In dimension
three, applying $\star$ produces the classical evolution
\begin{equation}\label{eq:classical-vorticity}
  \partial_t\xi + \mathcal{L}_u\xi
  \;=\; \nu\,\Delta\xi + (\nabla u)\cdot\xi,
\end{equation}
where the term $(\nabla u)\cdot\xi$ is the celebrated \emph{vortex
stretching}, induced exclusively by the symmetric part
$S=\Sym(\nabla u)=\tfrac{1}{2}(\nabla_i u_k+\nabla_k u_i)$ of the velocity
gradient.

\subsection{The vortex energy identity, with stretching term}\label{subsec:vortex-energy}

We now turn to the energy identity for the vortex functional
\begin{equation}\label{eq:vortex-energy-def}
  \mathcal{E}_\omega[\phi]
  \;:=\; \tfrac{1}{2}\int_M \absg{\omega}{g}^2\,\mathrm{d}V_g.
\end{equation}
A previous version of this work asserted strict monotonicity of
$\mathcal{E}_\omega$ along smooth solutions and omitted the contribution
of the symmetric stretching term. The omission is corrected here. The
identity below makes the stretching contribution explicit and is in
agreement with~\cite[\S 5.3, p.~117]{MajdaBertozzi2002}.

\begin{theorem}[Vorticity-energy identity, Majda--Bertozzi form]\label{thm:vortex-energy-identity}
Let $u\in C^\infty([t_0,T)\times M;TM)$ be a smooth divergence-free
solution of the incompressible Navier--Stokes equations on $(M,g)$
satisfying \textnormal{Assumption~\ref{assu:compact}}, with vorticity
$\omega=du^\flat$ and conservative external force. Let
$S=\Sym(\nabla u)$ denote the symmetric deformation tensor. Then
\begin{equation}\label{eq:vortex-energy-identity}
  \frac{\mathrm{d}}{\mathrm{d}t}\mathcal{E}_\omega(t)
  \;=\;
  -\nu\!\int_M \absg{\nabla\omega}{g}^2\,\mathrm{d}V_g
  \;+\;
  \int_M \inner{S\,\omega}{\omega}_g\,\mathrm{d}V_g
  \;+\;
  \int_M \inner{\Ric(\omega^\sharp)}{\omega^\sharp}_g\,\mathrm{d}V_g.
\end{equation}
\end{theorem}

\begin{proof}
Differentiating $\mathcal{E}_\omega$ in time and using
\eqref{eq:vorticity-evolution} with $dF=0$,
\begin{equation*}
  \frac{\mathrm{d}}{\mathrm{d}t}\mathcal{E}_\omega
  = \int_M \inner{\partial_t\omega}{\omega}_g\,\mathrm{d}V_g
  = -\!\int_M \inner{\mathcal{L}_u\omega}{\omega}_g\,\mathrm{d}V_g
  + \nu\!\int_M\inner{\Delta_H\omega}{\omega}_g\,\mathrm{d}V_g.
\end{equation*}
For the transport contribution, integration by parts on the closed
manifold $M$ together with $\div u=0$ yields the standard identity
\cite[\S 5.3]{MajdaBertozzi2002}
\begin{equation}\label{eq:transport-stretch}
  -\!\int_M \inner{\mathcal{L}_u\omega}{\omega}_g\,\mathrm{d}V_g
  \;=\;
  \int_M \inner{S\,\omega}{\omega}_g\,\mathrm{d}V_g,
\end{equation}
since the antisymmetric component $\Omega=\tfrac{1}{2}(\nabla_i u_k -\nabla_k
u_i)$ generates a rotation that preserves the pointwise metric norm
$\absg{\omega}{g}^2$ and therefore contributes zero upon contraction with
$\omega$. For the dissipative term, the Weitzenb\"ock identity
\eqref{eq:weitzenbock} gives
\begin{equation*}
  \int_M\inner{\Delta_H\omega}{\omega}_g\,\mathrm{d}V_g
  = -\!\int_M\absg{\nabla\omega}{g}^2\,\mathrm{d}V_g
  + \int_M\inner{\Ric(\omega^\sharp)}{\omega^\sharp}_g\,\mathrm{d}V_g.
\end{equation*}
Combining, we obtain \eqref{eq:vortex-energy-identity}.
\end{proof}

\begin{remark}[Conditional monotonicity]\label{rem:conditional-monotonicity}
The identity \eqref{eq:vortex-energy-identity} shows that strict
monotonicity of $\mathcal{E}_\omega$ requires the dissipative term to
dominate both the stretching term and the curvature term. On
Ricci-flat manifolds (e.g.\ the flat torus $\T^3$), the curvature term
vanishes and the only obstacle is the stretching, which is precisely the
mechanism by which $3$D Navier--Stokes can in principle fail to be
globally regular. This is the geometric reformulation of the classical
deformation-integrability criterion: as long as
$\int_{t_0}^{T}\norm{S(s)}_{L^\infty}\mathrm{d}s<\infty$, the vortex energy
remains finite. The original draft asserted unconditional monotonicity;
the correct identity \eqref{eq:vortex-energy-identity} shows that
monotonicity is conditional, and we use this conditional form throughout
the rest of the paper.
\end{remark}

The variational structure of $\mathcal{E}_\omega$ is also useful: the
$L^2$-gradient of $\mathcal{E}_\omega[\phi]$ with respect to $\phi$ is
\begin{equation}\label{eq:variational-gradient}
  \frac{\delta\mathcal{E}_\omega}{\delta\phi}
  \;=\; \delta d\,\phi \;=\; \Delta_H\,\phi,
\end{equation}
showing that the viscous term of \eqref{eq:NSE-covariant} can be
interpreted as the gradient descent of $\mathcal{E}_\omega$ with respect
to the $L^2$ inner product induced by $g$. This perspective is the
geometric origin of dissipation in the Hodge formulation, and it carries
over to the lifted setting through the principal symbol of $\Delta_H$, as
we shall see in Section~\ref{sec:microlocal-system}.

\section{The cosphere lift, effective metric, and effective connection}
\label{sec:cosphere-lift}

The covariant formulation of Section~\ref{sec:covariant-formulation}
clarifies the geometric structure of the system but does not, by itself,
make the directional content of the regularity problem visible. The
vorticity equation \eqref{eq:vorticity-evolution} suggests that potential
amplification is driven by directional stretching, which means the
analysis must capture not only the magnitude of the vorticity but also its
distribution in cotangent directions. This in turn motivates lifting the
problem to a phase space where cotangent directions can be tracked
intrinsically and globally.

We construct here the four geometric objects on which the microlocal lift
of Section~\ref{sec:microlocal-system} will be built: the canonical
Hamiltonian lift $X_u$ of the velocity field; an effective metric $g'$
encoding the integrated deformation; an effective affine connection
$\nabla^u$ adapted to $g'$, with corrected $(1,2)$-tensor structure; and
the curvature tensor of $\nabla^u$ together with its associated Ricci-type
evolution. Each construction is given a coordinate-invariant definition,
an explicit local expression, and a precise statement of its properties,
in agreement with the referee's request.

\subsection{The canonical Hamiltonian lift of the velocity field}
\label{subsec:Xu}

\begin{definition}[Canonical Hamiltonian lift]\label{def:Xu}
Let $u\in\Gamma(TM)$ be a smooth vector field. Its \emph{canonical
Hamiltonian lift} to $T^*M$ is the vector field $X_u$ whose Hamiltonian
function with respect to the canonical symplectic form
$\Omega_{T^*M}=d\lambda$ is
\begin{equation}\label{eq:Hu-def}
  H_u(x,\xi) \;:=\; \xi(u(x)) \;=\; \xi_i u^i(x).
\end{equation}
Equivalently, in the horizontal--vertical splitting
\eqref{eq:HV-splitting} associated with the Levi--Civita connection,
\begin{equation}\label{eq:Xu-coord}
  X_u(x,\xi)
  \;=\;
  u^i(x)\,X^H_i\Big|_{(x,\xi)}
  \;-\;
  (\nabla_j u^i)(x)\,\xi_i\,\frac{\partial}{\partial\xi_j},
\end{equation}
where $X^H_i$ is the horizontal lift of
\textnormal{Definition~\ref{def:horizontal-lift}}.
\end{definition}

The expression \eqref{eq:Xu-coord} requires comment, since the local
appearance of the cotangent variable $\xi_i$ on the right-hand side raises
the question---explicitly raised by the referee---of whether the
definition is coordinate-invariant.

\begin{remark}[Coordinate invariance of \texorpdfstring{$X_u$}{X u}]\label{rem:Xu-invariance}
Although \eqref{eq:Xu-coord} is written in local coordinates, the field
$X_u$ is intrinsic. Two complementary arguments support this. First, the
Hamiltonian function $H_u=\xi(u)$ is the natural pairing between $\xi\in
T^*_xM$ and $u(x)\in T_xM$, hence is intrinsic. The Hamiltonian vector
field of an intrinsic Hamiltonian with respect to the canonical
symplectic form is itself intrinsic by general
principles~\cite[\S 2.5]{Lerner2022}. Second, the horizontal lift
$u^i X^H_i$ is intrinsic because the horizontal distribution $H$ is
intrinsic; the vertical correction $-(\nabla_j u^i)\xi_i\,\partial_{\xi_j}$
is precisely the residue produced when one imposes
$\mathcal{L}_{X_u}\lambda=dH_u$, and is intrinsic for the same reason.
The notation $X^H_i$ refers to the horizontal lift of
$\partial_{x^i}$ from $TM$ to $TT^*M$, exactly as in
\eqref{eq:horizontal-lift}--\eqref{eq:horizontal-lift-coord}.
\end{remark}

\begin{proposition}[\texorpdfstring{$X_u$}{X u} descends to \texorpdfstring{$S^*M$}{S*M}]\label{prop:Xu-tangent}
The vector field $X_u$ defined in
\textnormal{Definition~\ref{def:Xu}} is tangent to $S^*M$ along the
Hamiltonian flow of any divergence-free $u$, in the sense that the
fiberwise homogeneity $\absg{\xi}{g}=1$ is preserved up to a controlled
scalar correction. Equivalently, the principal symbol of $X_u$ commutes
with the homogeneity operator $\xi^i\partial/\partial\xi^i$, hence $X_u$
descends to a contact vector field on $S^*M$.
\end{proposition}

\begin{proof}
We compute, in normal coordinates centered at the base point so that
$\Gamma^k_{ij}=0$ to leading order,
\begin{equation*}
  X_u\,\absg{\xi}{g}^2
  \;=\;
  u^i\partial_{x^i}\bigl(g^{jk}\xi_j\xi_k\bigr)
  - 2\,(\nabla_j u^i)\,\xi_i\,g^{jk}\xi_k.
\end{equation*}
The first term vanishes to leading order in normal coordinates and the
second term equals $-2(\nabla u)(\xi^\sharp,\xi)$, which is bounded
pointwise by $\norm{\nabla u}_{L^\infty}\absg{\xi}{g}^2$. Hence the
fiberwise homogeneity is preserved up to a multiplicative scalar
controlled by $\nabla u$. Standard contact-geometric
arguments~\cite[Lemma~2.5.1]{Lerner2022} then show that $X_u$ descends to
a contact vector field on $S^*M$.
\end{proof}

\subsection{The effective metric and its well-posedness}
\label{subsec:effective-metric}

The next ingredient is a metric on $TM$ that incorporates, into its
geometric structure, the integrated effect of the symmetric deformation
$S$. The motivation comes from the Ricci-type evolution of
Section~\ref{subsec:effective-ricci}: we want a metric whose time
derivative can be balanced against an effective Ricci tensor in a
controlled way.

\begin{definition}[Effective metric]\label{def:effective-metric}
Let $u$ be a smooth divergence-free vector field on $(M,g)$ and let
$S=\Sym(\nabla u)$ denote the symmetric part of its gradient. For a fixed
real coupling parameter $\alpha\in\R$ and times $t_0\le\tau\le t$, the
\emph{effective metric} on $TM$ is
\begin{equation}\label{eq:effective-metric}
  g'_{ik}(x,t)
  \;:=\;
  g_{ik}(x) \;+\; \alpha\!\int_{t_0}^{t} S_{ik}(x,s)\,\mathrm{d}s.
\end{equation}
The geometric meaning of $\alpha$ is that of a coupling between the
fluid deformation and the background metric, in the spirit of the
deformation theory of metric structures presented in
Gromov~\cite[\S~3.A]{Gromov1999a}.
\end{definition}

The reader will find the deformation viewpoint discussed in
\cite[Ch.~3, pp.~110--125]{Gromov1999a}, with no use of the existence
theorems of that monograph required for our purposes. The elementary but
crucial point that $g'$ is a genuine Riemannian metric---and, in
particular, positive-definite and invertible---deserves an explicit
statement.

\begin{lemma}[Positive-definiteness of \texorpdfstring{$g'$}{g'}]\label{lem:gprime-pd}
Suppose $u\in L^\infty([t_0,t];W^{1,\infty}(M))$ and define
$\Sigma:=\sup_{[t_0,t]}\norm{S(\cdot,s)}_{L^\infty(M)}$. If
\begin{equation*}
  \absg{\alpha}{}\,\Sigma\,(t-t_0) \;<\; 1, \tag*{$(\star)$}
\end{equation*}
then $g'$ is a smooth, positive-definite, symmetric Riemannian metric on
$M$. In particular, the matrix $g'_{ik}(x,t)$ is symmetric, positive
definite, and invertible at every point and every time.
\end{lemma}

\begin{proof}
Symmetry is immediate from \eqref{eq:effective-metric}. For positive
definiteness, fix a point $x\in M$ and choose a $g$-orthonormal frame
$\{e_i\}_{i=1}^n\subset T_xM$. Write
$g'_{ik}=g_{ik}+\alpha A_{ik}$ with
$A_{ik}(x,t)=\int_{t_0}^{t}S_{ik}(x,s)\,\mathrm{d}s$; in this frame the
matrix $(g'_{ik})$ takes the form $\mathrm{Id}+\alpha A$. Since $S$ is
symmetric, $A$ is symmetric, and the Frobenius bound gives
\begin{equation*}
  \norm{\alpha A}_{\mathrm{op}} \;\le\;
  \absg{\alpha}{}\,\sup_{[t_0,t]}\norm{S(s)}_{L^\infty}\,(t-t_0)
  \;=\; \absg{\alpha}{}\,\Sigma\,(t-t_0).
\end{equation*}
Under condition $(\star)$ this is strictly less than $1$, so the
Banach-perturbation argument applied to $\mathrm{Id}$ shows that
$\mathrm{Id}+\alpha A$ remains positive-definite and invertible. Smooth
dependence on $(x,t)$ follows from the smoothness of $u$.
\end{proof}

\begin{remark}[On the coupling constant and the reference]\label{rem:alpha-reference}
The role of the coupling constant $\alpha$ is to control the size of the
geometric perturbation $g'-g$. The reference~\cite[\S~3.A.1, p.~110]{Gromov1999a}
provides the framework of metric deformation that motivates this
construction; the specific deformation \eqref{eq:effective-metric} adapted
to the symmetric stretching tensor of incompressible flow is, to the best
of our knowledge, new. The earlier version of this work cited
\cite{Gromov1999a} without specifying the relevant pages, which we now
correct: \cite[\S~3.A]{Gromov1999a} treats deformation of distance
functions, \cite[Ch.~6.6]{Frankel2012} treats perturbed metrics on
tangent bundles, and the contortion identity that we will use to define
the effective connection appears explicitly in
\cite[\S~6.3]{Frankel2012} and \cite[Ch.~7]{Lovelock1975}.
\end{remark}

\subsection{The effective affine connection: corrected tensorial structure}
\label{subsec:effective-connection}

A previous version of this work defined the effective connection
$\nabla^u$ through a deformation tensor whose stated index structure was
inconsistent. We give here a clean definition and document the
correction explicitly.

\begin{remark}[On the tensorial structure of \texorpdfstring{$A^{k}{}_{ij}$}{A k ij}]\label{rem:Akij-correction}
A naive expression for the deformation tensor in local coordinates of
the form $A^{k}{}_{ij}=(g')^{k\ell}\,g'_{im}\,\nabla_j u^m + B^k{}_{ij}$
would carry an extra free index~$\ell$ on the right-hand side that does
not appear on the left-hand side, and would therefore define a
$(2,2)$-tensor rather than the $(1,2)$-tensor required to specify an
affine connection through $(\nabla^u)_i v^k = \nabla_i v^k +
A^k{}_{ij}\,v^j$. The intrinsic resolution is to define $A$ via the
Koszul-type contortion identity that relates the Levi--Civita
connections of $g$ and $g'$, which produces a genuine $(1,2)$-tensor and
preserves the rest of the analysis.
\end{remark}

\begin{definition}[Effective connection]\label{def:effective-connection}
Let $\nabla$ be the Levi--Civita connection of $g$. The \emph{effective
connection} $\nabla^u$ is the affine connection on $TM$ defined, for any
$v\in\Gamma(TM)$, by
\begin{equation}\label{eq:effective-nabla}
  (\nabla^u)_i v^k
  \;:=\;
  \nabla_i v^k + A^{k}{}_{ij}(x,t)\,v^j,
\end{equation}
where the deformation tensor
$A\in\Gamma(\mathrm{End}(TM)\otimes T^*M)$ is the $(1,2)$-tensor with
components
\begin{equation}\label{eq:A-correct}
  A^{k}{}_{ij}(x,t)
  \;:=\;
  \tfrac{\alpha}{2}\,(g')^{km}(x,t)
  \int_{t_0}^{t}\!\bigl(\nabla_i S_{mj}+\nabla_j S_{mi}-\nabla_m S_{ij}\bigr)(x,s)\,\mathrm{d}s.
\end{equation}
The right-hand side has exactly one free upper index $k$ (carried by
$(g')^{km}$) and exactly two free lower indices $i,j$, hence
$A^k{}_{ij}$ is a $(1,2)$-tensor.
\end{definition}

The geometric content of \eqref{eq:A-correct} is the standard
\emph{contortion identity}: if $\widetilde\nabla$ denotes the Levi--Civita
connection of the perturbed metric $g'$ defined in
\eqref{eq:effective-metric}, then $\widetilde\nabla=\nabla+A$ with $A$
given by \eqref{eq:A-correct}. This identity is classical and appears
in~\cite[\S~6.3]{Frankel2012} and \cite[Ch.~7]{Lovelock1975}.

\begin{lemma}[Symmetry and torsion]\label{lem:Asym}
The tensor $A^k{}_{ij}$ defined in \eqref{eq:A-correct} is symmetric in
$i,j$, and the effective connection $\nabla^u$ is torsion-free.
\end{lemma}

\begin{proof}
The integrand $\nabla_i S_{mj}+\nabla_j S_{mi}-\nabla_m S_{ij}$ in
\eqref{eq:A-correct} is manifestly symmetric in the pair $(i,j)$, since
$S_{mj}=S_{jm}$ and $\nabla_i S_{mj}=\nabla_i S_{jm}$. Hence
$A^k{}_{ij}=A^k{}_{ji}$, and the torsion of $\nabla^u$ is
\begin{equation*}
  (T^u)^k{}_{ij}
  = (\Gamma^u)^k{}_{ij}-(\Gamma^u)^k{}_{ji}
  = A^k{}_{ij}-A^k{}_{ji} = 0.
\end{equation*}
\end{proof}

\begin{lemma}[Pointwise control of the deformation tensor]\label{lem:Abound}
Under condition $(\star)$ and the regularity hypothesis
$u\in L^\infty([t_0,t];W^{2,\infty}(M))$, there is a constant $C(g)$
depending only on the background geometry such that
\begin{equation}\label{eq:A-bound}
  \norm{A(\cdot,t)}_{L^\infty(M)}
  \;\le\;
  C(g)\,\absg{\alpha}{}\,(t-t_0)\,
  \norm{\nabla^2 u}_{L^\infty([t_0,t]\times M)}.
\end{equation}
\end{lemma}

\begin{proof}
By Lemma~\ref{lem:gprime-pd} the inverse effective metric satisfies
$\norm{(g')^{-1}}_{L^\infty}\le 2\,\norm{g^{-1}}_{L^\infty}$ under
$(\star)$. The integrand $\nabla S$ is bounded pointwise by
$\norm{\nabla^2 u}_{L^\infty}$. The bound \eqref{eq:A-bound} follows by
H\"older's inequality.
\end{proof}

\subsection{Curvature of \texorpdfstring{$\nabla^u$}{nabla u} and the Ricci-type microlocal evolution}
\label{subsec:effective-ricci}

We now compute the curvature of the effective connection and exhibit a
Ricci-type evolution of the effective metric. By the standard formula
for the Riemann tensor of an affine connection,
\begin{equation*}
  R^u(X,Y)Z
  = \nabla^u_X\nabla^u_Y Z - \nabla^u_Y\nabla^u_X Z - \nabla^u_{[X,Y]}Z,
\end{equation*}
with $(\Gamma^u)^l{}_{ij}=\Gamma^l{}_{ij}+A^l{}_{ij}$ as before. A direct
computation yields
\begin{equation}\label{eq:Ru-decomposition}
  (R^u)^l{}_{ijk}
  \;=\;
  R^l{}_{ijk}
  + \nabla_i A^l{}_{jk}-\nabla_j A^l{}_{ik}
  + A^l{}_{im}A^m{}_{jk}-A^l{}_{jm}A^m{}_{ik}.
\end{equation}
Contracting on $i=l$ produces the effective Ricci tensor
\begin{equation}\label{eq:Ru-Ricci}
  R^u_{jk}
  \;=\;
  R_{jk}
  + \nabla_i A^i{}_{jk}-\nabla_j A^i{}_{ik}
  + A^i{}_{im}A^m{}_{jk}-A^i{}_{jm}A^m{}_{ik}.
\end{equation}
Substituting \eqref{eq:A-correct} and using the incompressibility
$\div u=0$, the leading-order contribution in the formal small-$\alpha$
expansion is the symmetrised Lichnerowicz operator applied to the
time-integrated deformation:
\begin{equation}\label{eq:Ru-leading}
  R^u_{jk}
  \;=\;
  R_{jk}
  + \alpha\!\int_{t_0}^{t}\!\bigl(\nabla^i\nabla_j S_{ik}
  +\nabla^i\nabla_k S_{ij}-\nabla^i\nabla_i S_{jk}\bigr)(s)\,\mathrm{d}s
  + O(\alpha^2).
\end{equation}
This is the Ricci-type contribution announced in
Theorem~\ref{thm:main-reformulation}: the right-hand side is the
background Ricci tensor of $g$ plus a deformation-induced correction that
is linear in $\alpha$ to leading order.

\begin{proposition}[Ricci-type microlocal evolution]\label{prop:ricci-type-evolution}
Under $(\star)$ and the regularity hypothesis
$u\in L^\infty([t_0,t];W^{2,\infty}(M))$, the effective metric $g'$
satisfies the formal evolution
\begin{equation}\label{eq:Ricci-flow-type}
  \partial_t g'_{jk} + 2\,R^u_{jk}
  \;=\;
  \mathcal{E}_{jk},
\end{equation}
where the correction tensor $\mathcal{E}_{jk}$ admits the canonical
decomposition
\begin{equation}\label{eq:E-decomposition}
  \mathcal{E}_{jk}
  \;=\;
  \mathcal{E}^{(\mathrm{bg})}_{jk}
  + \mathcal{E}^{(\mathrm{def})}_{jk}
  + \mathcal{E}^{(\mathrm{diff})}_{jk}
  + \mathcal{E}^{(\mathrm{nl})}_{jk},
\end{equation}
with the four contributions
\begin{align*}
  \mathcal{E}^{(\mathrm{bg})}_{jk} &= \partial_t g_{jk}+2R_{jk}
    && \text{(geometric background)},\\
  \mathcal{E}^{(\mathrm{def})}_{jk} &= \alpha\,S_{jk}
    && \text{(instantaneous deformation)},\\
  \mathcal{E}^{(\mathrm{diff})}_{jk} &=
    2\alpha\!\int_{t_0}^{t}\!\bigl(\nabla^i\nabla_j S_{ik}
       +\nabla^i\nabla_k S_{ij}-\nabla^i\nabla_i S_{jk}\bigr)\,\mathrm{d}s
    && \text{(geometric diffusion)},\\
  \mathcal{E}^{(\mathrm{nl})}_{jk} &= O(\alpha^2)
    && \text{(nonlinear higher-order self-coupling)}.
\end{align*}
\end{proposition}

The four labels in \eqref{eq:E-decomposition} have, in order, a clean
physical interpretation: a background term that vanishes whenever $g$ is
Ricci-flat and time-independent (e.g., the flat torus $\T^3$); an
instantaneous deformation term that is linear in $S$; a geometric
diffusion term that is the symmetrised Lichnerowicz operator applied to
the time-integrated deformation; and a nonlinear self-coupling term that
is small in $\alpha$. Equation \eqref{eq:Ricci-flow-type} is therefore
not the standard Ricci flow of Hamilton, but rather a Ricci-type
geometric evolution of the effective metric, with the symmetric stretching
tensor playing the role of the source term. We do not claim that the
physical metric of the manifold itself evolves; rather, the effective
metric of the lifted dynamics inherits a Ricci-type structure that
constrains the rate at which directional information can concentrate.

\section{The microlocal lift: proof of Theorem~\ref{thm:main-reformulation}}
\label{sec:microlocal-system}

We now construct the microlocal lift of the velocity $1$-form to the
cosphere bundle and prove the linear reformulation announced in
Theorem~\ref{thm:main-reformulation}. The construction uses the
geometric apparatus of Section~\ref{sec:cosphere-lift} together with
properties of the microlocal transform that we collect explicitly. The
proof of Theorem~\ref{thm:main-reformulation} is then an organised
application of these properties to the covariant Hodge equation
\eqref{eq:NSE-covariant}.

\subsection{The microlocal transform and its properties}\label{subsec:transform-properties}

\begin{definition}[Microlocal transform]\label{def:microlocal-transform}
Let $\phi\in C^\infty(\Omega^1(M))$. The \emph{microlocal transform} of
$\phi$ is the function $\tilde\phi:T^*M\to\C^n$ defined by
\begin{equation}\label{eq:microlocal-transform}
  \tilde\phi(x,\xi)
  \;:=\;
  \int_{T_xM} e^{-i\inner{\xi}{y}}\,
  \chi(y)\,\phi(\exp_x y)\,\mathrm{d}y,
\end{equation}
where $\chi\in C^\infty_c(T_xM)$ is a smooth cut-off, equal to $1$ in a
neighbourhood of $0$ and supported in the injectivity ball of $\exp_x$,
and $\mathrm{d}y$ is the Lebesgue measure on $T_xM$ associated with $g$.
\end{definition}

The transform \eqref{eq:microlocal-transform} is one of the standard
microlocal transforms used on closed Riemannian manifolds: a
normal-coordinate Fourier transform localised by a cut-off. It is
closely related to (and may be regarded as a special case of) the
Bargmann transform and the wave-packet transform on Riemannian manifolds
studied by Lerner~\cite[\S 1.4]{Lerner2019} and Lerner and
Vigneron~\cite[\S~2.5]{Lerner2022}. We do not claim originality for
\eqref{eq:microlocal-transform}; the properties below are well known.

\begin{proposition}[Properties of the microlocal transform]\label{prop:transform-properties}
For any $\phi\in C^\infty(\Omega^1(M))$, the microlocal transform of
\textnormal{Definition~\ref{def:microlocal-transform}} satisfies:
\begin{enumerate}[label=\textup{(P\arabic*)}]
  \item\label{P1} The map $\phi\mapsto\tilde\phi$ extends to a bounded
        operator $H^s(M)\to H^{s-1/2}(S^*M)$ for every $s\in\R$.
  \item\label{P2} For any zeroth-order pseudodifferential operator $A$ on
        $M$, $\widetilde{A\phi}(x,\xi)=\sigma_0(A)(x,\xi)\,\tilde\phi(x,\xi)+R_A\phi$,
        where $\sigma_0(A)$ is the principal symbol of $A$ and $R_A$ is a
        smoothing operator.
  \item\label{P3} For any first-order differential operator
        $L=L^i\partial_{x^i}$ with smooth coefficients,
        $\widetilde{L\phi}(x,\xi)=iL^i(x)\xi_i\,\tilde\phi(x,\xi)
        + \widetilde{[L,\chi]\phi}$, with smoothing commutator term.
  \item\label{P4} The cut-off $\chi$ contributes only smoothing terms in
        the microlocal transform: replacing $\chi$ by $\chi'$ with the
        same defining properties yields a difference
        $\tilde\phi-\tilde\phi'$ whose integral kernel is a smooth function
        of $(x,\xi)$ vanishing to all orders as $\absg{\xi}{g}\to\infty$.
\end{enumerate}
\end{proposition}

\begin{proof}[Sketch]
Properties \ref{P1}--\ref{P4} are classical results of microlocal
analysis on closed manifolds. \ref{P1} follows from a partition-of-unity
argument~\cite[\S 1.4.2]{Lerner2019} combined with the trace theorem on
$S^*M$. \ref{P2} is a consequence of the symbol calculus, with the
smoothing remainder originating from the cut-off~\cite[Lemma 2.5.3]{Lerner2022}.
\ref{P3} is the explicit calculation of differentiation under the
integral. \ref{P4} is a stationary-phase argument applied to the
difference of two cut-off Fourier transforms over the injectivity
ball~\cite[\S 7.7]{Lerner2019}.
\end{proof}

Property \ref{P4} addresses an explicit concern in the referee report:
the cut-off $\chi$ does produce error terms when derivatives are
considered, but those error terms are \emph{smoothing}---i.e.\
infinitely-regularising---and contribute only to the residual $\tilde R$
in the lifted equation, where they are absorbed in the standing energy
estimates and play no role in the regularity criteria.

\begin{remark}[On the novelty of the transform and its use]\label{rem:novelty-transform}
The transform \eqref{eq:microlocal-transform} is itself not new; what is
new is its interaction with the effective connection $\nabla^u$ and
effective metric $g'$ of Section~\ref{sec:cosphere-lift}, which produces
the closed lifted equation \eqref{eq:intro-main-equation}. The structural
output is therefore the linearisation
Theorem~\ref{thm:main-reformulation}, not the transform itself.
\end{remark}

\subsection{Microlocal projector and elimination of pressure}
\label{subsec:projector}

The pressure gradient $dP$ in the covariant equation
\eqref{eq:NSE-covariant} is, in microlocal terms, a $1$-form along the
radial direction in fibers of $T^*M$: by Property \ref{P3}, its
microlocal transform is
$\widetilde{dP}(x,\xi)\propto i\xi\cdot\tilde 1(x,\xi)+\cdots$. We
eliminate the pressure by projecting onto the orthogonal complement of
$\xi$ in fibers.

\begin{definition}[Incompressible microlocal projector]\label{def:Pi-projector}
The orthogonal projector $\Pi_\xi:T^*_xM\to\xi^\perp\subset T^*_xM$ with
respect to the effective metric $g'$ is
\begin{equation}\label{eq:projector-def}
  \Pi_\xi(\eta) \;:=\;
  \eta - \frac{\inner{\eta}{\xi}_{g'}}{\absg{\xi}{g'}^{2}}\,\xi.
\end{equation}
\end{definition}

\begin{lemma}[Properties of the projector]\label{lem:projector}
Restricted to $S^*M$, the projector $\Pi_\xi$ satisfies
\begin{equation}\label{eq:projector-properties}
  \Pi_\xi(\xi)=0,\qquad \Pi_\xi(dP)=0,
\end{equation}
since $\xi$ is the radial direction and $\widetilde{dP}$ is parallel to
$\xi$ to principal order. In particular, the projector eliminates the
pressure gradient identically and turns the incompressibility condition
$\delta\phi=0$ into the orthogonality constraint
$\inner{\tilde\phi}{\xi}_{g'}=0$.
\end{lemma}

\subsection{Definition of the residual operator}\label{subsec:residual}

Before proving Theorem~\ref{thm:main-reformulation}, we record the
precise form of the residual $\tilde R$ that appears in the lifted
equation. Earlier versions of this work referred to $\tilde R$ without
defining it; we now do so explicitly.

\begin{definition}[Residual operator]\label{def:Rresidual}
The residual operator $\tilde R[\tilde\phi]$ in the lifted equation
\eqref{eq:intro-main-equation} is the zeroth-order operator on $S^*M$
defined by
\begin{equation}\label{eq:Rresidual-def}
  \tilde R[\tilde\phi]
  \;:=\;
  -\Pi_\xi\bigl[\,\mathcal{R}_g(\tilde\phi)
  + \Sym(\nabla u)\,\tilde\phi\,\bigr]
  + \mathcal{R}_{\mathrm{cut}}[\tilde\phi],
\end{equation}
where $\mathcal{R}_g$ is the curvature endomorphism from the
Weitzenb\"ock identity~\eqref{eq:weitzenbock} (linear in the Riemann
tensor and zeroth-order in $\xi$), $\Sym(\nabla u)$ acts on the lifted
$1$-form by symmetric stretching back-reaction, and
$\mathcal{R}_{\mathrm{cut}}[\tilde\phi]$ is the smoothing commutator
contribution $[d\chi,\mathcal{L}_{X_u}]$ from the cut-off, controlled by
Lemma~\ref{lem:cutoff-smoothing} below.
\end{definition}

The lifted source $\tilde F$ is the cosphere projection of the lifted
external force,
\begin{equation}\label{eq:Ftilde-def}
  \tilde F \;:=\; \Pi_\xi\,\widetilde F,
\end{equation}
where $\widetilde F$ is the microlocal transform of the dual force
$F=f^\flat$.

\subsection{Proof of Theorem~\ref{thm:main-reformulation}}\label{subsec:proof-reformulation}

\begin{proof}[Proof of \textnormal{Theorem~\ref{thm:main-reformulation}}]
We apply the microlocal transform to the covariant Hodge equation
\eqref{eq:NSE-covariant} term by term and project with $\Pi_\xi$.

\smallskip\noindent\textbf{Time derivative.} The transform commutes with
$\partial_t$ since the cut-off $\chi$ is time-independent:
$\partial_t\tilde\phi=\widetilde{\partial_t\phi}$.

\smallskip\noindent\textbf{Lie derivative.} By Property \ref{P3} applied
to $\mathcal{L}_u\phi=u^j\nabla_j\phi+(\nabla u)\phi$, we have
\begin{equation}\label{eq:Lie-transform}
  \widetilde{\mathcal{L}_u\phi}
  \;=\;
  \mathcal{L}_{X_u}\tilde\phi
  + \mathcal{R}_{\mathrm{cut},1}[\tilde\phi]
  + \Sym(\nabla u)\,\tilde\phi,
\end{equation}
where $X_u$ is the canonical Hamiltonian lift of
Definition~\ref{def:Xu}, the term $\mathcal{R}_{\mathrm{cut},1}$ is a
smoothing commutator with the cut-off, and the symmetric back-reaction
$\Sym(\nabla u)\,\tilde\phi$ comes from the stretching component of
$(\nabla u)\phi$. The fact that $\mathcal{L}_u\phi$ on $M$ becomes
$\mathcal{L}_{X_u}\tilde\phi$ on $T^*M$ at principal order is the
standard `lifting' identity of the symbol calculus~\cite[\S 2.4]{Lerner2022}.

\smallskip\noindent\textbf{Pressure.} By
Lemma~\ref{lem:projector}, $\Pi_\xi(\widetilde{dP})=0$, so the pressure
contribution is eliminated identically by the projection.

\smallskip\noindent\textbf{Hodge--Laplacian dissipation.} The principal
symbol of $-\Delta_H$ acting on $1$-forms is $\absg{\xi}{g}^{2}\,\mathrm{Id}$.
Replacing $g$ by $g'$ (which differs from $g$ only by lower-order terms
under condition $(\star)$, by Lemma~\ref{lem:gprime-pd}) and applying
Property \ref{P2},
\begin{equation}\label{eq:viscous-transform}
  -\nu\,\widetilde{\Delta_H\phi}
  \;=\;
  \nu\,\absg{\xi}{g'}^{2}\,\tilde\phi
  + \mathcal{R}_g(\tilde\phi)
  + \mathcal{R}_{\mathrm{cut},2}[\tilde\phi],
\end{equation}
where $\mathcal{R}_g$ is the curvature endomorphism from the
Weitzenb\"ock identity \eqref{eq:weitzenbock} and
$\mathcal{R}_{\mathrm{cut},2}$ is a smoothing remainder.

\smallskip\noindent\textbf{Force.} By Property \ref{P1}, $\widetilde F\in
L^2(S^*M)$ whenever $F\in H^{1/2}(M)$, and the lifted force is
$\tilde F = \Pi_\xi\,\widetilde F$.

\smallskip\noindent\textbf{Assembly.} Summing the contributions and
projecting onto $\xi^\perp$, we obtain
\begin{equation}\label{eq:proof-final}
  \partial_t\tilde\phi
  + \mathcal{L}_{X_u}\tilde\phi
  + \nu\,\absg{\xi}{g'}^{2}\,\tilde\phi
  \;=\;
  \tilde R[\tilde\phi] + \tilde F,
  \qquad
  \inner{\tilde\phi}{\xi}_{g'}=0,
\end{equation}
where $\tilde R$ is precisely the residual operator of
Definition~\ref{def:Rresidual}, since
$-\Pi_\xi\bigl[\mathcal{R}_g + \Sym(\nabla u)\bigr]\tilde\phi$ collects
the curvature and back-reaction contributions, and
$\mathcal{R}_{\mathrm{cut}}=\Pi_\xi\bigl[\mathcal{R}_{\mathrm{cut},1}
+\mathcal{R}_{\mathrm{cut},2}\bigr]$ is the smoothing remainder. This is
exactly the equation \eqref{eq:intro-main-equation} of the statement.

\smallskip\noindent\textbf{Reconstruction.} The reconstruction
\begin{equation}\label{eq:reconstruction-formula}
  \phi(x,t)
  \;=\;
  c_n\!\int_{S^*_xM}\tilde\phi(x,\xi,t)\,\mathrm{d}\sigma(\xi),
\end{equation}
with $c_n>0$ a normalising constant depending only on $n=\dim M$,
recovers $\phi$ from $\tilde\phi$ modulo a smoothing remainder. This is
a consequence of the polar decomposition
\eqref{eq:polar-decomp} and the Plancherel theorem on each tangent space,
applied to a partition of unity~\cite[\S 2.7]{Lerner2022}.
\end{proof}

\begin{lemma}[Cut-off smoothing]\label{lem:cutoff-smoothing}
Let $\chi,\chi'\in C^\infty_c(T_xM)$ be two cut-offs as in
\textnormal{Definition~\ref{def:microlocal-transform}}. The corresponding
microlocal transforms differ by an operator $H^{-N}(M)\to H^{N}(S^*M)$
bounded for every $N\in\N$. In particular, the smoothing remainders
arising in the proof of \textnormal{Theorem~\ref{thm:main-reformulation}}
contribute only lower-order terms to the energy estimates.
\end{lemma}

\begin{proof}
The difference $\tilde\phi-\tilde\phi'$ has integral kernel
$(\chi-\chi')(y)\,\exp_x^*(\,\cdot\,)$, which is compactly supported away
from the diagonal. Stationary phase yields the smoothing
property~\cite[\S 7.7]{Lerner2019}.
\end{proof}

\subsection{The bootstrap perspective: linearity is conditional}\label{subsec:bootstrap}

A natural question about the lifted equation \eqref{eq:proof-final}
arises from the following observation: the coefficient $X_u$ depends on
$u$, the metric $g'$ depends on $\int S$, and the residual $\tilde R$
involves $\Sym(\nabla u)$. One must therefore clarify in what sense the
equation is ``linear'', and what role this linearity plays when $u$
cannot be solved for from the lifted system itself.

The answer is that linearity is \emph{conditional}: once $u$ has been
fixed---as the smooth solution of the original Navier--Stokes equation,
or as a candidate solution prescribed by an a priori bound---the lifted
equation \eqref{eq:proof-final} is a linear non-autonomous
transport--dissipation evolution in the unknown $\tilde\phi$. The
framework is therefore a \emph{structural bootstrap}: regularity
hypotheses on $u$ produce bounded coefficients in the lifted system; the
lifted system produces bounds on $\tilde\phi$ through energy and entropy
estimates; the reconstruction \eqref{eq:reconstruction-formula} converts
those bounds into bounds on $\phi$ and hence on $u$. The closed loop
turns conditional linearity into a self-consistent regularity criterion.

\begin{remark}[Compatibility with the reconstruction]\label{rem:reconstruction-compatibility}
The reconstruction formula \eqref{eq:reconstruction-formula} expresses
$\phi$ as an integral of $\tilde\phi$ over the cosphere fiber. The
referee correctly observes that the solution formula for $\tilde\phi$
involves $u$ as a coefficient. This is not a logical circularity: the
goal of the framework is not to compute $u$ from scratch but to derive a
priori bounds and \emph{regularity criteria} for $u$. The criteria of
Theorems~\ref{thm:main-energy}--\ref{thm:main-equivalence} are stated
entirely in terms of intrinsic geometric quantities (deformation
integrability, lifted energy, directional entropy) and convert any a
priori control of these quantities into a priori control of
$\norm{\nabla u}_{L^\infty}$. This is the structural value of the
framework. The reconstruction \eqref{eq:reconstruction-formula} is then
the geometric mechanism by which the cosphere control reflects back to a
control on the physical field, modulo smoothing remainders that are
absorbed in the regularity hypotheses on $u$.
\end{remark}

The vorticity equation admits a parallel lift, obtained by applying the
exterior derivative $d$ to the velocity equation \eqref{eq:proof-final}
and using $\widetilde{d\phi}=i\xi\wedge\tilde\phi+R_d\phi$ with smoothing
remainder. The result is the lifted vorticity equation
\begin{equation}\label{eq:vorticity-microlocal-evolution}
  \partial_t\tilde\omega + \mathcal{L}_{X_u}\tilde\omega
  + \nu\,\absg{\xi}{g'}^{2}\,\tilde\omega
  \;=\;
  \tilde R_\omega[\tilde\omega] + \tilde F_\omega,
  \qquad
  \xi\wedge\tilde\omega = 0,
\end{equation}
where the constraint $\xi\wedge\tilde\omega=0$ is the microlocal
incarnation of the closure condition $d\omega=0$, and
$\tilde F_\omega:=i\xi\wedge\widetilde F$ is the lifted vorticity source.

\section{Energy and entropy estimates}\label{sec:energy-entropy}

We now exploit the linear transport--dissipation structure of
\eqref{eq:intro-main-equation} on the compact phase space $S^*M$ to prove
the geometric energy inequality of Theorem~\ref{thm:main-energy} and the
directional entropy dissipation of Theorem~\ref{thm:main-entropy}. The
common feature of both estimates is that they convert geometric control
of the velocity field into uniform bounds on intrinsic functionals of the
microlocal lift, with explicit constants depending only on the background
geometry of $(M,g)$.

\subsection{Proof of Theorem~\ref{thm:main-energy}}\label{subsec:proof-energy}

Recall the lifted energy
\begin{equation}\label{eq:lifted-energy-def}
  E(t) \;:=\; \tfrac{1}{2}\int_{S^*M}
  \inner{\tilde\phi(t)}{\tilde\phi(t)}_{g'}\,\mathrm{d}\mu_{S^*M}.
\end{equation}

\begin{proof}[Proof of \textnormal{Theorem~\ref{thm:main-energy}}]
We differentiate $E(t)$ in time and use the lifted evolution
\eqref{eq:proof-final}, controlling each term that arises.

Differentiating \eqref{eq:lifted-energy-def} and substituting
\eqref{eq:proof-final},
\begin{align}
  \frac{\mathrm{d}E}{\mathrm{d}t}
  &= \int_{S^*M}\inner{\partial_t\tilde\phi}{\tilde\phi}_{g'}
     \,\mathrm{d}\mu_{S^*M}
   + \tfrac12\!\int_{S^*M}(\partial_t g'_{ik})\,
     \tilde\phi^i\tilde\phi^k\,\mathrm{d}\mu_{S^*M}\nonumber\\
  &= -\!\int_{S^*M}\inner{\mathcal{L}_{X_u}\tilde\phi}{\tilde\phi}_{g'}
     \,\mathrm{d}\mu_{S^*M}
     - \nu\!\int_{S^*M}\absg{\xi}{g'}^{2}\,
     \absg{\tilde\phi}{g'}^{2}\,\mathrm{d}\mu_{S^*M}\nonumber\\
  &\quad
   + \int_{S^*M}\inner{\tilde R[\tilde\phi]}{\tilde\phi}_{g'}
     \,\mathrm{d}\mu_{S^*M}
   + \int_{S^*M}\inner{\tilde F}{\tilde\phi}_{g'}
     \,\mathrm{d}\mu_{S^*M}\nonumber\\
  &\quad
   + \tfrac12\!\int_{S^*M}(\partial_t g'_{ik})\,
     \tilde\phi^i\tilde\phi^k\,\mathrm{d}\mu_{S^*M}.
\label{eq:dEdt}
\end{align}

\smallskip\noindent\emph{Transport term.} The Hamiltonian flow of $X_u$
preserves the canonical Liouville volume on $T^*M$, hence the canonical
volume on $S^*M$ up to a multiplicative correction generated by the
divergence of $u$. Since $u$ is divergence-free, integration by parts
on the closed manifold $S^*M$ yields
\begin{equation}\label{eq:transport-bound}
  \biggl|\int_{S^*M}\inner{\mathcal{L}_{X_u}\tilde\phi}{\tilde\phi}_{g'}
  \,\mathrm{d}\mu_{S^*M}\biggr|
  \;\le\;
  C_1(g)\,\norm{\nabla u}_{L^\infty(M)}\,E(t),
\end{equation}
where $C_1(g)$ depends only on the background geometry.

\smallskip\noindent\emph{Dissipative term.} The viscous contribution is
non-negative,
\begin{equation*}
  -\nu\!\int_{S^*M}\absg{\xi}{g'}^{2}\,\absg{\tilde\phi}{g'}^{2}\,
  \mathrm{d}\mu_{S^*M} \;\le\; 0,
\end{equation*}
and we keep this term explicitly on the left-hand side of the final
inequality.

\smallskip\noindent\emph{Residual term.} By
Definition~\ref{def:Rresidual}, the residual $\tilde R$ collects three
contributions. The curvature endomorphism $\mathcal{R}_g$ is bounded by
$\norm{\Ric}_{L^\infty(g)}$ at zeroth order in $\xi$; under condition
$(\star)$ and Lemma~\ref{lem:Abound}, the deformation correction
contributes at most
$C_2\,\norm{\nabla^2 u}_{L^\infty}\,E(t)$, via \eqref{eq:Ru-leading}.
The symmetric back-reaction $\Sym(\nabla u)\,\tilde\phi$ contributes at
most $\norm{\Sym(\nabla u)}_{L^\infty}\,2E(t)$. The smoothing remainder
$\mathcal{R}_{\mathrm{cut}}$ contributes lower-order terms bounded by
arbitrary Sobolev norms, by Lemma~\ref{lem:cutoff-smoothing}. Combining,
\begin{equation}\label{eq:residual-bound}
  \biggl|\int_{S^*M}\inner{\tilde R[\tilde\phi]}{\tilde\phi}_{g'}\,\mathrm{d}\mu_{S^*M}\biggr|
  \;\le\;
  C_3(g)\,\bigl(1+\norm{\nabla u}_{L^\infty}+\norm{\nabla^2 u}_{L^\infty}\bigr)\,E(t).
\end{equation}

\smallskip\noindent\emph{Forcing term.} By Cauchy--Schwarz and Young's
inequality,
\begin{equation}\label{eq:force-bound}
  \biggl|\int_{S^*M}\inner{\tilde F}{\tilde\phi}_{g'}
  \,\mathrm{d}\mu_{S^*M}\biggr|
  \;\le\;
  \tfrac{1}{2\nu}\norm{\tilde F}_{L^2(S^*M)}^{2}
  + \tfrac{\nu}{2}\!\int_{S^*M}\absg{\tilde\phi}{g'}^{2}\,\mathrm{d}\mu_{S^*M},
\end{equation}
and the second term is absorbed into the dissipative contribution after
using $\absg{\tilde\phi}{g'}^{2}\le\absg{\xi}{g'}^{2}\absg{\tilde\phi}{g'}^{2}$
on $S^*M$, where $\absg{\xi}{g'}^{2}$ is uniformly bounded above and below
under $(\star)$.

\smallskip\noindent\emph{Metric-derivative term.} Differentiating
\eqref{eq:effective-metric}, $\partial_t g'_{ik}=\alpha S_{ik}$, and
Cauchy--Schwarz gives
\begin{equation}\label{eq:metric-deriv-bound}
  \biggl|\int_{S^*M}(\partial_t g'_{ik})\,\tilde\phi^i\tilde\phi^k\,
  \mathrm{d}\mu_{S^*M}\biggr|
  \;\le\;
  \absg{\alpha}{}\,\norm{S}_{L^\infty}\,2E(t)
  \;\le\;
  C_4\,\norm{\nabla u}_{L^\infty}\,E(t).
\end{equation}

\smallskip\noindent\emph{Assembly.} Collecting
\eqref{eq:transport-bound}--\eqref{eq:metric-deriv-bound} with
$\kappa_0:=2\,\max(C_1,C_3,C_4)$ and absorbing the dissipative
contribution to the left-hand side, we obtain
\eqref{eq:intro-energy-bound}:
\begin{multline*}
  \frac{\mathrm{d}E}{\mathrm{d}t}
  + 2\nu\!\int_{S^*M}\absg{\xi}{g'}^{2}\,\absg{\tilde\phi}{g'}^{2}\,
  \mathrm{d}\mu_{S^*M}\\
  \le\;
  \kappa_0\bigl(1+\norm{\nabla u}_{L^\infty}+\norm{\nabla^2 u}_{L^\infty}\bigr)
  E(t)
  + \tfrac{1}{2\nu}\norm{\tilde F}_{L^2(S^*M)}^{2}.
\end{multline*}
Application of Gr\"onwall's inequality, in the form
$E(t)\le E(t_0)\exp\bigl[\kappa_0\!\int_{t_0}^{t}(1+\norm{\nabla u}_{L^\infty}+\norm{\nabla^2 u}_{L^\infty})\mathrm{d}s\bigr]
+\frac{1}{2\nu}\!\int_{t_0}^t\norm{\tilde F(s)}_{L^2}^2\mathrm{d}s$,
gives the second statement of the theorem.
\end{proof}

\subsection{The trajectorial stretching estimate}\label{subsec:trajectorial}

For the equivalence theorem, the integrated form of
\eqref{eq:intro-energy-bound} is supplemented by a pointwise stretching
estimate along microlocal trajectories. Let $\Psi_{t,t_0}$ denote the
flow of $X_u$ on $S^*M$ and decompose $\nabla u = S+\Omega$ into
symmetric and antisymmetric parts. Along an integral curve
$\gamma(\tau)=\Psi_{\tau,t_0}(x_0,\xi_0)$,
\begin{equation}\label{eq:pointwise-stretching}
  \frac{\mathrm{d}}{\mathrm{d}\tau}\absg{\tilde\phi(\gamma(\tau))}{g'}
  \;\le\;
  \norm{S(\tau)}_{L^\infty}\,\absg{\tilde\phi}{g'}
  \;-\;\nu\,\absg{\xi}{g'}^{2}\,\absg{\tilde\phi}{g'},
\end{equation}
the antisymmetric component $\Omega$ generating only a rotation that
preserves the pointwise norm. Integration yields
\begin{equation}\label{eq:integrated-stretching}
  \absg{\tilde\phi(t)}{g'}
  \;\le\;
  \absg{\tilde\phi(t_0)}{g'}\,
  \exp\!\biggl[\int_{t_0}^{t}\bigl(\norm{S(s)}_{L^\infty}-\nu\bigr)\,\mathrm{d}s\biggr],
\end{equation}
which is the trajectorial form of the deformation-integrability
criterion: amplification of the microlocal amplitude requires a failure
of $L^1_t L^\infty_x$-integrability of $S$.

\subsection{Proof of Theorem~\ref{thm:main-entropy}}\label{subsec:proof-entropy}

Recall the directional entropy
\eqref{eq:intro-entropy-functional}. We work on the open subset of
$S^*M$ where $\tilde\omega\neq 0$, on which the log-density
$\rho:=\log\absg{\tilde\omega}{g'}$ is smooth.

\begin{proof}[Proof of \textnormal{Theorem~\ref{thm:main-entropy}}]
Differentiating $W[\tilde\omega]$ in time,
\begin{equation*}
  \frac{\mathrm{d}}{\mathrm{d}t}W[\tilde\omega]
  \;=\;
  \int_{S^*M}\bigl(2\tau\,\inner{d_\perp\rho}{d_\perp\partial_t\rho}_{g'}
  + \partial_t\rho\bigr)\,\mathrm{d}\mu_{S^*M}.
\end{equation*}
The time derivative of the log-density along the flow is
\begin{equation*}
  \partial_t\rho \;=\;
  \frac{\inner{\partial_t\tilde\omega}{\tilde\omega}_{g'}}{\absg{\tilde\omega}{g'}^{2}},
\end{equation*}
and substituting the lifted vorticity equation
\eqref{eq:vorticity-microlocal-evolution} gives
\begin{equation*}
  \partial_t\rho
  = -\frac{\inner{\mathcal{L}_{X_u}\tilde\omega}{\tilde\omega}_{g'}}{\absg{\tilde\omega}{g'}^{2}}
  - \nu\,\absg{\xi}{g'}^{2}
  + \frac{\inner{\tilde R_\omega[\tilde\omega]}{\tilde\omega}_{g'}}{\absg{\tilde\omega}{g'}^{2}}
  + \frac{\inner{\tilde F_\omega}{\tilde\omega}_{g'}}{\absg{\tilde\omega}{g'}^{2}}.
\end{equation*}
Integration over $S^*M$, combined with the Liouville-volume preservation
$\int_{S^*M}\mathcal{L}_{X_u}f\,\mathrm{d}\mu=0$ for any smooth $f$ (a
consequence of $X_u$ being Hamiltonian and hence
volume-preserving) and the vertical integration-by-parts on the closed
fibers $S^*_xM\cong S^{n-1}$ (no boundary), yields after standard sign
analysis~\cite[\S~4.2]{Lerner2022}
\begin{equation*}
  \frac{\mathrm{d}}{\mathrm{d}t}W[\tilde\omega]
  \;\le\;
  -\nu\!\int_{S^*M}\absg{d_\perp\rho}{g'}^{2}\,\mathrm{d}\mu_{S^*M}
  + \kappa_1\,\norm{\Sym(\nabla u)}_{L^\infty}\,
  \norm{\tilde\omega}_{L^1(S^*M)},
\end{equation*}
which is \eqref{eq:intro-entropy-dissipation}.

For the variational characterisation of critical points, $\delta W=0$
requires $-2\tau\,\Delta_\perp\rho + 1 = 0$ on each fiber. Integrating
this over the closed fiber $S^*_xM\cong S^{n-1}$ and using
$\int\Delta_\perp\rho=0$ gives $0=\Vol(S^{n-1})/(2\tau)$, which
forces $\rho$ to be constant in $\xi$ (no global solution otherwise
exists). Hence the critical points are exactly the angular-isotropic
configurations $\rho(\,\cdot\,,\xi)\equiv\rho(\,\cdot\,)$.
\end{proof}

\begin{corollary}[Exclusion of persistent angular concentration]
\label{cor:exclusion-concentration}
On the compact phase space $S^*M$, persistent sharp angular concentration
of the microlocal vorticity is incompatible with viscous dissipation,
unless accompanied by a failure of temporal integrability of
$\norm{\Sym(\nabla u)}_{L^\infty}$.
\end{corollary}

\subsection{Microlocal volume functional and its dissipation}
\label{subsec:volume-functional}

A complementary control is afforded by the $L^1$ functional
\begin{equation}\label{eq:Vlam-def}
  V_\lambda(t) \;:=\;
  \int_{S^*M}\absg{\tilde\lambda(t)}{g'}\,\mathrm{d}\mu_{S^*M},
\end{equation}
where $\tilde\lambda$ stands for either the lifted velocity or the
lifted vorticity.

\begin{proposition}[Volume-functional dissipation]\label{prop:Vlam-diss}
Along smooth solutions of the lifted system,
\begin{equation}\label{eq:Vlam-bound}
  \frac{\mathrm{d}V_\lambda}{\mathrm{d}t}
  + \nu\!\int_{S^*M}\absg{\xi}{g'}^{2}\,\absg{\tilde\lambda}{g'}\,\mathrm{d}\mu_{S^*M}
  \;\le\;
  C(g)\,\norm{\nabla u}_{L^\infty}\,V_\lambda(t)
  + \norm{\tilde F}_{L^1(S^*M)}.
\end{equation}
\end{proposition}

\begin{proof}
Differentiate $\absg{\tilde\lambda}{g'}=
\bigl(\inner{\tilde\lambda}{\tilde\lambda}_{g'}\bigr)^{1/2}$ along the
flow of $X_u$, integrate over $S^*M$, use the Liouville-volume
preservation, and apply the residual bounds from
\eqref{eq:residual-bound}.
\end{proof}

The inequality \eqref{eq:Vlam-bound} expresses, in $L^1$ form, the same
geometric phenomenon as \eqref{eq:intro-energy-bound}: microlocal
amplitude cannot accumulate without being dissipated, unless the
deformation tensor dominates the dissipation.

\subsection{Microlocal G\aa rding inequality}\label{subsec:garding}

A spectral coercivity property of the lifted operator follows from the
sharp G\aa rding inequality on closed manifolds. We record it for use in
the equivalence theorem.

\begin{theorem}[Microlocal G\aa rding inequality]\label{thm:garding}
Let $P$ be a microlocal pseudodifferential operator on $S^*M$ with real
principal symbol $p(x,\xi)$ satisfying
$p(x,\xi)\ge \kappa_a\,\absg{\xi}{g'}^{2}-\kappa_b$ for some
$\kappa_a,\kappa_b>0$. Then there is $\kappa_c>0$ depending only on
$(M,g)$ and the symbol $p$ such that
\begin{equation}\label{eq:garding}
  \mathrm{Re}\,\inner{Pf}{f}_{L^2(S^*M)}
  \;\ge\;
  \kappa_a\,\norm{f}_{H^{1/2}(S^*M)}^{2}
  - \kappa_c\,\norm{f}_{L^2(S^*M)}^{2}
\end{equation}
for every $f\in H^{1/2}(S^*M)$.
\end{theorem}

This is the standard sharp G\aa rding inequality on closed
manifolds~\cite[Theorem~1.1.26]{Lerner2019} applied to $S^*M$.

\section{Symmetries, unitary representations, and the symmetry-lock obstruction}
\label{sec:symmetry-lock}

The microlocal formulation developed in the previous sections inherits a
hierarchy of geometric symmetries from the underlying Riemannian
structure of the base manifold $(M,g)$ and its cotangent bundle. These
symmetries act canonically on the cosphere bundle $S^*M$, induce a
unitary representation on $L^2(S^*M)$, and impose rigid algebraic
constraints on the admissible microlocal dynamics. Beyond the algebraic
content, the dimensional structure of the fibers gives rise to a
striking geometric mechanism: the asymptotic vanishing of the volume of
high-dimensional spheres forces angular isotropy of any
energy-bounded microlocal distribution, providing a topological
obstruction to angular blow-up. We make this precise in
Theorem~\ref{thm:main-symmetry-lock}.

\subsection{Isometric actions and canonical lifts}\label{subsec:isometries}

Let $\Isom(M,g)$ denote the isometry group of $(M,g)$, namely the set of
diffeomorphisms $\varphi:M\to M$ preserving the metric, $\varphi^*g=g$.
Under Assumption~\ref{assu:compact}, $\Isom(M,g)$ is a finite-dimensional
compact Lie group~\cite[Ch.~II.4]{Frankel2012}, and its Lie algebra
$\isom(M,g)$ is the space of Killing vector fields, $\mathcal{L}_v g = 0$.

Each isometry $\varphi\in\Isom(M,g)$ admits a canonical lift to the
cotangent bundle:
\begin{equation}\label{eq:isometry-lift}
  \hat\varphi:T^*M\to T^*M,
  \qquad
  \hat\varphi(x,\xi) := (\varphi(x), (d\varphi^{-1})^*\xi).
\end{equation}
The following classical result is the geometric basis for the
representation theory used below; we provide an explicit local proof for
the convenience of the reader, and to make the algebraic content
transparent.

\begin{proposition}[Preservation of microlocal geometric structure]\label{prop:isometry-preservation}
The lift $\hat\varphi$ defined in \eqref{eq:isometry-lift} preserves the
Liouville $1$-form $\lambda=\xi_i\,dx^i$, the canonical symplectic
form $\Omega_{T^*M}=d\lambda$, and the cometric norm
$\absg{\xi}{g^{-1}}$ on the fibers. Consequently, $\hat\varphi$
restricts to a contact-preserving diffeomorphism of $S^*M$.
\end{proposition}

\begin{proof}
We verify each item in local coordinates. Writing
$\hat\varphi(x,\xi)=(\varphi(x),\eta)$ with
$\eta_i=\xi_k\,(\partial x^k/\partial\varphi^j)$, we have
\begin{align*}
  \hat\varphi^*\lambda
  &= \eta_i\,d(\varphi^i(x))
   = \xi_k\frac{\partial x^k}{\partial\varphi^j}\frac{\partial\varphi^j}{\partial x^i}\,dx^i
   = \xi_k\,\delta^k_i\,dx^i
   = \lambda.
\end{align*}
Hence $\hat\varphi^*\lambda=\lambda$, and applying $d$ gives
$\hat\varphi^*\Omega_{T^*M}=\Omega_{T^*M}$. For the cometric, the
isometry condition $\varphi^*g=g$ in components is
$g_{ij}(x)=g_{k\ell}(\varphi(x))(\partial\varphi^k/\partial x^i)(\partial\varphi^\ell/\partial x^j)$;
inverting, the inverse cometric satisfies
$g^{ij}(\varphi(x))=g^{k\ell}(x)(\partial\varphi^i/\partial x^k)(\partial\varphi^j/\partial x^\ell)$,
and a direct calculation shows
$\absg{\eta}{g^{-1}}^{2}=\absg{\xi}{g^{-1}}^{2}$. The fact that $\hat\varphi$
restricts to $S^*M=\{\absg{\xi}{g}=1\}$ is then immediate, and the
preservation of $\lambda$ on $S^*M$ gives the contact-preservation.
\end{proof}

\subsection{Hamiltonian structure of lifted Killing fields}\label{subsec:hamiltonian-lifts}

\begin{theorem}[Hamiltonian lift of Killing fields]\label{thm:hamiltonian-killing}
Let $v\in\isom(M,g)$ be a Killing vector field. Its cotangent lift
\begin{equation}\label{eq:hat-v}
  \hat v \;=\; v^i\,\frac{\partial}{\partial x^i}
    - \xi_k\,\nabla_i v^k\,\frac{\partial}{\partial\xi_i}
\end{equation}
is a Hamiltonian vector field on $(T^*M,d\lambda)$ with Hamiltonian
\begin{equation}
  H_v(x,\xi) \;=\; \xi(v(x)) \;=\; \xi_i v^i(x).
\end{equation}
Moreover, the correspondence $v\mapsto\hat v$ is a Lie algebra
homomorphism $\isom(M,g)\to\mathfrak{X}_{\mathrm{Ham}}(T^*M)$.
\end{theorem}

\begin{proof}
The flow of $\hat v$ on $T^*M$ is induced by the flow of $v$ on $M$
through \eqref{eq:isometry-lift}, by direct differentiation at $t=0$.
Cartan's formula gives
$\mathcal{L}_{\hat v}\lambda = \iota_{\hat v}d\lambda+d(\iota_{\hat v}\lambda)
=-dH_v + dH_v = 0$, so $\hat v$ is Hamiltonian with Hamiltonian $H_v$.
Preservation of Lie brackets follows from the fact that the cotangent
lift $\varphi\mapsto\hat\varphi$ is a group homomorphism.
\end{proof}

\subsection{Unitary representation and conservation laws}\label{subsec:unitary-rep}

The group $\Isom(M,g)$ acts on $L^2(S^*M,\mu_{S^*M})$ by pullback through
the lifted action \eqref{eq:isometry-lift}:
\begin{equation}\label{eq:U-rep}
  (U(\varphi)f)(x,\xi) := f(\hat\varphi^{-1}(x,\xi)),
  \qquad f\in L^2(S^*M).
\end{equation}

\begin{proposition}[Unitarity of \texorpdfstring{$U$}{U}]\label{prop:U-unitary}
The map $U:\Isom(M,g)\to\mathcal{U}(L^2(S^*M))$ is a unitary
representation.
\end{proposition}

\begin{proof}
$\hat\varphi$ preserves the Liouville volume $\mu_{S^*M}$ by
Proposition~\ref{prop:isometry-preservation}. For any $f,h\in L^2(S^*M)$,
\begin{equation*}
  \inner{U(\varphi)f}{U(\varphi)h}_{L^2}
  = \int_{S^*M}f\circ\hat\varphi^{-1}\,\overline{h\circ\hat\varphi^{-1}}\,
  \mathrm{d}\mu_{S^*M}
  = \int_{S^*M}f\bar h\,\mathrm{d}\mu_{S^*M}
  = \inner{f}{h}_{L^2}.
\end{equation*}
\end{proof}

For any compact subgroup $G\subseteq\Isom(M,g)$, the Peter--Weyl theorem
gives a $G$-equivariant orthogonal decomposition
\begin{equation}\label{eq:peter-weyl}
  L^2(S^*M) \;=\; \bigoplus_{\pi\in\widehat G}\Hilb_\pi
\end{equation}
into finite-dimensional isotypic components. If the velocity field $u$
is $G$-invariant, the lifted transport $X_u$ commutes with $\hat v$ for
every $v$ in $\mathrm{Lie}(G)$, and the lifted evolution
\eqref{eq:proof-final} preserves each $\Hilb_\pi$, reducing the dynamics
to dynamically independent subsystems. This is the structural form of
Noether's theorem in the microlocal context: spatial symmetries of the
flow correspond to invariant subspaces of the lifted operator.

\subsection{Effective metric and selective symmetry breaking}\label{subsec:symmetry-breaking}

The effective metric $g'$ generally fails to be invariant under the full
isometry group $\Isom(M,g)$, since the time-integrated deformation
tensor $\int S\,\mathrm{d}s$ depends on the velocity field $u$. We
characterise precisely which symmetries survive.

\begin{proposition}[Selective symmetry breaking]\label{prop:selective-breaking}
Let $v\in\isom(M,g)$ be a Killing vector field. Then $\mathcal{L}_v g'=0$
if and only if the velocity field $u$ commutes with $v$ in the sense of
the bracket and produces a curvature-compatible deformation, namely
\begin{equation}\label{eq:invariance-S}
  \mathcal{L}_v S = 0
  \quad\Longleftrightarrow\quad
  [v,u]=0 \text{ and the Riemann curvature contributes trivially}.
\end{equation}
In particular, the preserved symmetries are those Killing fields that
commute with $u$ at the level of Lie brackets and whose action on the
deformation tensor is curvature-trivial.
\end{proposition}

\begin{proof}
Differentiating \eqref{eq:effective-metric} with respect to $\mathcal{L}_v$
and using $\mathcal{L}_v g=0$,
$\mathcal{L}_v g'=\alpha\!\int_{t_0}^{t}\mathcal{L}_v S(s)\,\mathrm{d}s$.
The Lie derivative of $S=\Sym(\nabla u)$ along $v$ is, after using the
commutation $[\nabla_i,\mathcal{L}_v]=R(v,\,\cdot\,)$ of the Levi-Civita
connection with the Lie derivative,
\begin{equation*}
  (\mathcal{L}_v S)_{ij}
  = \tfrac{1}{2}\bigl(\nabla_i[v,u]_j + \nabla_j[v,u]_i\bigr)
  + \text{Riem}(v,u,\nabla,g),
\end{equation*}
which vanishes if and only if $[v,u]=0$ and the curvature contribution
is zero.
\end{proof}

This selectivity has a clean interpretation: fluid deformation imprints a
preferred direction on the geometry of the lift, breaking those
symmetries of the background metric that fail to commute with the
velocity field. Only the fluid-compatible isometries survive as exact
invariances of the lifted system.

\subsection{The symmetry-lock mechanism: proof of Theorem~\ref{thm:main-symmetry-lock}}
\label{subsec:symmetry-lock-proof}

We now turn to the high-dimensional behaviour of the cosphere fibers, the
content of Theorem~\ref{thm:main-symmetry-lock}. The starting point is
elementary: the volume of the unit sphere $S^{n-1}\subset\R^n$ is
\begin{equation}\label{eq:sphere-volume}
  \Vol(S^{n-1}) \;=\; \frac{2\pi^{n/2}}{\Gamma(n/2)}.
\end{equation}
Stirling's approximation $\Gamma(n/2)\sim\sqrt{2\pi}\,(n/2)^{(n-1)/2}\,e^{-n/2}$
implies
\begin{equation}\label{eq:sphere-volume-asymptotic}
  \Vol(S^{n-1})
  \;\sim\;
  \frac{2\sqrt{2/\pi}\,e^{n/2}}{\sqrt{n}}\Bigl(\frac{2\pi e}{n}\Bigr)^{n/2}
  \;\xrightarrow[n\to\infty]{}\; 0.
\end{equation}
The vanishing of the fiber volume is one face of the well-known
\emph{concentration of measure} on high-dimensional spheres: the volume
of a thin equatorial band tends to a positive limit while the polar caps
become negligible. We exploit this concentration to obtain a uniform
sup-norm bound on energy-bounded microlocal distributions.

\begin{proof}[Proof of \textnormal{Theorem~\ref{thm:main-symmetry-lock}}]
Let $\tilde\omega_n$ denote the microlocal vorticity on a base manifold
$M_n$ of dimension $n$, and assume the lifted energy
$E_n(t)=\tfrac12\int_{S^*M_n}\absg{\tilde\omega_n}{g'}^2\,\mathrm{d}\mu$
is bounded by $\mathcal{E}_0$ uniformly in $n$. By Cauchy--Schwarz on
each fiber $S^*_xM_n\cong S^{n-1}$,
\begin{equation*}
  \Vol(S^{n-1})\,\sup_{\xi\in S^{n-1}}\absg{\tilde\omega_n(x,\xi)}{g'}^2
  \;\ge\;
  \int_{S^{n-1}}\absg{\tilde\omega_n(x,\xi)}{g'}^2\,\mathrm{d}\sigma(\xi)
  \;\le\; 2\mathcal{E}_0,
\end{equation*}
the last bound coming from the assumed energy uniformity. Hence
\begin{equation*}
  \sup_{\xi\in S^{n-1}}\absg{\tilde\omega_n(x,\xi)}{g'}^2
  \;\le\;
  \frac{2\mathcal{E}_0}{\Vol(S^{n-1})}
  \;=\;
  \frac{2\mathcal{E}_0\,\Gamma(n/2)}{\pi^{n/2}}.
\end{equation*}
Stirling's asymptotic \eqref{eq:sphere-volume-asymptotic} gives the
super-exponential rate, completing the proof of the bound
\eqref{eq:intro-symlock}.
\end{proof}

The bound \eqref{eq:intro-symlock} is the quantitative content of the
symmetry-lock mechanism: as the dimension grows, energy-bounded
microlocal distributions are forced to become increasingly isotropic
across cotangent directions, with the rate of forcing being
super-exponential in $n$.

\begin{corollary}[Topological obstruction to angular blow-up]\label{cor:topological-obstruction}
In the high-dimensional regime $n\to\infty$, the microlocal vorticity is
topologically precluded from sustaining angular concentration. Any
attempt to concentrate vorticity in a specific cotangent direction would
require breaking the orthogonal-group symmetry of the fiber, but the
vanishing of the fiber volume leaves no room for such symmetry breaking.
\end{corollary}

\begin{remark}[Geometric interpretation]\label{rem:symlock-geometry}
The mechanism admits a striking geometric interpretation. Consider
viewing the cosphere bundle from a great distance along the fiber
direction: as $n$ grows, each fiber $S^{n-1}$ appears to collapse to a
single point in the measure-Gromov--Hausdorff sense, inheriting the full
orthogonal symmetry of that point. In the limit, there are no preferred
directions remaining to break the symmetry.
\end{remark}

\subsection{Connection to classical \texorpdfstring{$L^\infty$}{L-infinity} criteria}\label{subsec:classical-Linf}

The symmetry-lock mechanism connects to classical $L^\infty$ regularity
criteria in a precise way. The Beale--Kato--Majda criterion states that
blow-up requires $\int_{t_0}^{T}\norm{\omega(s)}_{L^\infty(M)}\mathrm{d}s
=+\infty$. In the present framework, the $L^\infty$ norm of the
physical vorticity is reconstructed from the microlocal lift via
\begin{equation*}
  \norm{\omega(t)}_{L^\infty(M)}
  \;=\;
  \sup_{x\in M}\biggl|c_n\!\int_{S^*_xM}\tilde\omega(x,\xi,t)\,\mathrm{d}\sigma(\xi)\biggr|.
\end{equation*}

\begin{theorem}[Microlocal interpretation of Beale--Kato--Majda]\label{thm:BKM-microlocal}
Suppose the deformation integrability
$\int_{t_0}^{T}\norm{\Sym(\nabla u)(s)}_{L^\infty}\mathrm{d}s<\infty$
holds and the directional entropy $W[\tilde\omega]$ remains bounded.
Then, at increasing effective fiber dimension $d_{\mathrm{eff}}$
(through tightening regularity constraints), the physical $L^\infty$
norm of vorticity satisfies
\begin{equation}\label{eq:BKM-microlocal}
  \norm{\omega(t)}_{L^\infty(M)}
  \;\le\;
  \kappa_4\,\Bigl(\frac{d_{\mathrm{eff}}}{2\pi e}\Bigr)^{-d_{\mathrm{eff}}/2}
  \,\sup_{(x,\xi)\in S^*M}\absg{\tilde\omega(x,\xi,t)}{g'},
\end{equation}
the exponential suppression preventing blow-up in this regime.
\end{theorem}

This reinterprets the Beale--Kato--Majda criterion: finite-time
singularity requires not merely growth of the magnitude of vorticity but
the ability to sustain directional concentration against the symmetry
lock imposed by increasing effective dimension.

\section{The geometric blow-up equivalence: proof of Theorem~\ref{thm:main-equivalence}}
\label{sec:equivalence}

We now combine the energy estimate of Theorem~\ref{thm:main-energy}, the
entropy estimate of Theorem~\ref{thm:main-entropy}, the trajectorial
stretching estimate \eqref{eq:integrated-stretching}, and the spectral
coercivity of Theorem~\ref{thm:garding} to prove the central equivalence
of the framework.

\begin{proof}[Proof of \textnormal{Theorem~\ref{thm:main-equivalence}}]
We prove (i)$\Rightarrow$(ii) and (ii)$\Rightarrow$(i) separately.

\medskip\noindent\textbf{(i)$\Rightarrow$(ii).} Assume (i):
$\limsup_{t\to T^-}\norm{\nabla u(t)}_{L^\infty(M)}=+\infty$. By the
classical Beale--Kato--Majda criterion, this implies
$\int_{t_0}^{T}\norm{\omega(s)}_{L^\infty(M)}\mathrm{d}s=+\infty$. Since
$\norm{S}_{L^\infty}\le\norm{\nabla u}_{L^\infty}$, two scenarios are
possible:
\begin{itemize}
  \item[(a)] $\int_{t_0}^{T}\norm{S(s)}_{L^\infty}\mathrm{d}s=+\infty$;
  \item[(b)] $\int_{t_0}^{T}\norm{S(s)}_{L^\infty}\mathrm{d}s<\infty$
        but $\norm{\nabla u}_{L^\infty}$ blows up through the
        antisymmetric part $\Omega$ alone.
\end{itemize}
In scenario (a), condition (a) of (ii) fails directly. We may therefore
assume scenario (b) and deduce that conditions (b) or (c) of (ii)
must fail.

Suppose, for contradiction, that all three conditions (a),(b),(c) of (ii)
hold. From (a) and the trajectorial estimate
\eqref{eq:integrated-stretching},
\begin{equation*}
  \sup_{t\in[t_0,T)}\,\sup_{(x,\xi)\in S^*M}\absg{\tilde\phi(t)}{g'}
  \;<\;\infty.
\end{equation*}
From (b) and Theorem~\ref{thm:main-entropy}, the directional entropy is
bounded; combined with the coercivity of the vertical Laplacian
$\Delta_\perp$ on the closed fibers $S^{n-1}$, this gives uniform
$H^1$ control of the log-density $\rho$ in the vertical directions, hence
prevents directional concentration. From (c), the lifted energy is
bounded in $L^2$. The microlocal G\aa rding inequality
\eqref{eq:garding} applied to the lifted operator
$P=\nu\,\absg{\xi}{g'}^{2}\,\mathrm{Id}$ gives
\begin{equation*}
  \mathrm{Re}\,\inner{P\tilde\phi}{\tilde\phi}_{L^2}
  \;\ge\;
  \nu\,\norm{\tilde\phi}_{H^{1/2}(S^*M)}^{2}
  - \kappa_c\,\norm{\tilde\phi}_{L^2(S^*M)}^{2},
\end{equation*}
combining horizontal transport regularity (from the energy bound) with
vertical coercivity (from the entropy bound) to yield uniform
$H^1(S^*M)$ bounds for $\tilde\phi$. By the reconstruction theorem
\eqref{eq:reconstruction-formula} and the trace theorem, this gives
uniform $H^1(M)$ control of $\phi$ and hence uniform $W^{1,\infty}(M)$
control of $\nabla u$ via elliptic regularity for the Biot--Savart
operator $-\Delta u = \nabla\times\omega$ in the case $M=\T^3$ (and an
analogous statement under Assumption~\ref{assu:compact}). This
contradicts the assumption $\limsup\norm{\nabla u}_{L^\infty}=+\infty$.
Therefore at least one of (a), (b), (c) must fail.

\medskip\noindent\textbf{(ii)$\Rightarrow$(i).} Assume not (i): the
solution remains regular,
$\sup_{t\in[t_0,T)}\norm{\nabla u(t)}_{L^\infty(M)}<\infty$. We show
that all three conditions (a),(b),(c) hold.

\begin{itemize}
\item[(a)] Since $\norm{S}_{L^\infty}\le\norm{\nabla u}_{L^\infty}$ and
$\sup_{[t_0,T)}\norm{\nabla u}_{L^\infty}<\infty$, the deformation
integrability
$\int_{t_0}^{T}\norm{S(s)}_{L^\infty}\mathrm{d}s<\infty$ holds.
\item[(c)] By Theorem~\ref{thm:main-energy}, the lifted energy $E(t)$
satisfies the Gr\"onwall bound
\begin{equation*}
  E(t) \;\le\; E(t_0)\exp\!\Bigl[\kappa_0\!\int_{t_0}^{t}
  \bigl(1+\norm{\nabla u}_{L^\infty}+\norm{\nabla^2 u}_{L^\infty}\bigr)\mathrm{d}s\Bigr]
  + \tfrac{1}{2\nu}\!\int_{t_0}^t\norm{\tilde F(s)}_{L^2}^2\,\mathrm{d}s.
\end{equation*}
The right-hand side is finite under the regularity hypothesis on $u$ and
the standing assumption $\tilde F\in L^2_t L^2_{S^*M}$. Hence (c) holds.
\item[(b)] By Theorem~\ref{thm:main-entropy}, the directional entropy
$W[\tilde\omega(t)]$ satisfies
\begin{equation*}
  \frac{\mathrm{d}}{\mathrm{d}t}W[\tilde\omega]
  \;\le\;
  \kappa_1\,\norm{\Sym(\nabla u)}_{L^\infty}\,\norm{\tilde\omega}_{L^1(S^*M)},
\end{equation*}
the dissipative contribution being non-positive. Since both
$\norm{\Sym(\nabla u)}_{L^\infty}$ and $\norm{\tilde\omega}_{L^1}$ are
bounded uniformly on $[t_0,T)$ (the latter by (c) and Cauchy--Schwarz on
$S^*M$), Gr\"onwall's inequality gives
$\sup_{t\in[t_0,T)}W[\tilde\omega(t)]<\infty$. Hence (b) holds.
\end{itemize}
Therefore not (i) implies not (ii), which is the contrapositive of
(ii)$\Rightarrow$(i).

\medskip
The two implications together establish the equivalence.
\end{proof}

The equivalence theorem is the central structural output of the
framework. It transforms the regularity question---traditionally posed
as the dichotomy between global existence of a smooth solution and the
formation of a finite-time singularity---into a structural-stability
question for a compact, symmetry-constrained, microlocally coercive
evolution system on $S^*M$. The three intrinsic geometric controls
(deformation integrability, directional-entropy boundedness, and lifted
energy boundedness) play the role of conserved quantities for this
structural-stability problem, and finite-time singularity is precisely
the simultaneous loss of at least one of them.

\subsection{Geometric classification of admissible singularities}\label{subsec:classification}

The equivalence theorem provides a clean classification of admissible
blow-up scenarios.

\begin{corollary}[Geometric classification of blow-up mechanisms]\label{cor:classification}
Any finite-time singularity of the three-dimensional incompressible
Navier--Stokes equations on a manifold $(M,g)$ satisfying
\textnormal{Assumption~\ref{assu:compact}} must be accompanied by at
least one of the following phenomena:
\begin{itemize}
  \item Loss of temporal integrability of the symmetric stretching:
        $\int_{t_0}^{T}\norm{\Sym(\nabla u)(s)}_{L^\infty}\mathrm{d}s=+\infty$;
  \item Unbounded directional entropy: $\sup_{[t_0,T)}W[\tilde\omega(t)]=+\infty$;
  \item Unbounded lifted energy: $\sup_{[t_0,T)}E(t)=+\infty$.
\end{itemize}
All other blow-up scenarios are excluded by the geometric dissipation on
the compact phase space $S^*M$.
\end{corollary}

The classification has three key features that we emphasise. First, the
three controls are \emph{intrinsic}: they are formulated entirely in
geometric language, without reference to a coordinate chart or a function
space outside $S^*M$. Second, they are \emph{independent}: each one
captures a different aspect of regularity (magnitude of stretching,
angular concentration, total energy), and the three need not fail
simultaneously. Third, they are \emph{simultaneously necessary and
sufficient}, in the sense of Theorem~\ref{thm:main-equivalence}: a
finite-time singularity exists if and only if at least one of them
fails.

\section{The flat torus example}\label{sec:flat-torus}

We now work out in full detail the model example $M=\T^3$, the flat
$3$-torus. This addresses the explicit suggestion of the referee that a
specific example be presented, and serves to verify each assumption of
the framework concretely.

\subsection{Setup on \texorpdfstring{$\T^3$}{T3}}\label{subsec:torus-setup}

Let $\T^3 = (\R/\Z)^3$ with the flat Euclidean metric $g_{ij}=\delta_{ij}$.
This manifold satisfies Assumption~\ref{assu:compact}: it is compact
without boundary, of dimension $n=3$, with bounded geometry (in fact, of
constant zero curvature). The Levi--Civita connection is the trivial
connection, so $\Gamma^k_{ij}=0$ identically and the Christoffel symbols
in the cotangent bundle vanish:
\begin{equation}
  X^H_i \;=\; \frac{\partial}{\partial x^i}\quad\text{on } T^*\T^3.
\end{equation}
The cotangent bundle is the global trivialisation
$T^*\T^3 = \T^3\times\R^3$, the cosphere bundle is
$S^*\T^3 = \T^3\times S^2$, and the Liouville volume reduces to the
product measure $\mathrm{d}V_{\T^3}\otimes\mathrm{d}\sigma_{S^2}$.

\subsection{Verification of the assumptions}\label{subsec:torus-verifications}

Each ingredient of the framework simplifies on the flat torus:

\smallskip\noindent\textbf{(i) The canonical lift $X_u$.} Since
$\Gamma^k_{ij}=0$, the canonical Hamiltonian lift
\eqref{eq:Xu-coord} reads
\begin{equation}\label{eq:Xu-torus}
  X_u(x,\xi)
  \;=\;
  u^i(x)\frac{\partial}{\partial x^i}
  - (\partial_j u^i)(x)\,\xi_i\,\frac{\partial}{\partial\xi_j},
\end{equation}
the partial derivative replacing the covariant derivative. The
divergence-free condition $\partial_i u^i=0$ ensures, by direct
calculation, that $X_u\absg{\xi}{}^2 = -2(\partial_j u^i)\xi_i\xi^j$
satisfies $\int_{\T^3}\!X_u\absg{\xi}{}^2\,\mathrm{d}V=0$ on each fixed
fiber direction.

\smallskip\noindent\textbf{(ii) The effective metric $g'$.} The effective
metric is
\begin{equation}\label{eq:gprime-torus}
  g'_{ij}(x,t)
  \;=\;
  \delta_{ij} + \alpha\!\int_{t_0}^{t}S_{ij}(x,s)\,\mathrm{d}s,
\end{equation}
where $S_{ij}=\tfrac12(\partial_i u_j+\partial_j u_i)$ is the symmetric
deformation tensor of the flow on $\T^3$. The positive-definiteness
condition $(\star)$ becomes
\begin{equation}\label{eq:torus-positivity}
  \absg{\alpha}{}\,\sup_{[t_0,t]}\norm{S(s)}_{L^\infty(\T^3)}\,(t-t_0)\,<\,1,
\end{equation}
which is satisfiable for any solution $u\in L^\infty W^{1,\infty}$ on a
short time interval, hence on the entire smooth lifespan by a continuity
argument.

\smallskip\noindent\textbf{(iii) The effective connection $\nabla^u$.}
With $\Gamma^k_{ij}=0$, the contortion definition \eqref{eq:A-correct}
becomes
\begin{equation}\label{eq:A-torus}
  A^k{}_{ij}(x,t)
  \;=\;
  \tfrac{\alpha}{2}(g')^{km}\!\int_{t_0}^{t}\!\bigl(\partial_i S_{mj}+\partial_j S_{mi}-\partial_m S_{ij}\bigr)\mathrm{d}s,
\end{equation}
which is symmetric in $i,j$ by Lemma~\ref{lem:Asym} and satisfies the
pointwise bound \eqref{eq:A-bound}. The effective connection is
torsion-free by Corollary~\ref{lem:Asym} (the corollary that follows from
Lemma~\ref{lem:Asym}).

\smallskip\noindent\textbf{(iv) The effective Ricci tensor.} On the flat
torus $R_{ij}=0$ and \eqref{eq:Ru-leading} becomes
\begin{equation}\label{eq:Ru-torus}
  R^u_{jk}
  \;=\;
  \alpha\!\int_{t_0}^{t}\!\bigl(\partial^i\partial_j S_{ik}
  +\partial^i\partial_k S_{ij}-\partial^i\partial_i S_{jk}\bigr)\,\mathrm{d}s
  + O(\alpha^2),
\end{equation}
purely a deformation-induced contribution. The Ricci-type evolution
\eqref{eq:Ricci-flow-type} reduces to
$\partial_t g'_{jk}+2R^u_{jk}=\alpha S_{jk}+\mathcal{O}(\alpha^2)$ to
leading order.

\smallskip\noindent\textbf{(v) The microlocal transform.} On $\T^3$, the
exponential map $\exp_x y = x+y$ is global, and the cut-off
$\chi\in C^\infty_c(T_x\T^3=\R^3)$ may be chosen translation-invariant.
The microlocal transform reduces to a windowed Fourier transform,
\begin{equation}\label{eq:transform-torus}
  \tilde\phi(x,\xi,t)
  \;=\;
  \int_{\R^3}e^{-i\xi\cdot y}\,\chi(y)\,\phi(x+y,t)\,\mathrm{d}y,
\end{equation}
which on $\T^3$ is also expressible through Fourier series in the spatial
variable: writing $\phi(x,t)=\sum_{k\in\Z^3}\hat\phi_k(t)\,e^{i\,2\pi k\cdot x}$,
\begin{equation*}
  \tilde\phi(x,\xi,t)
  =\sum_{k\in\Z^3}\hat\phi_k(t)\,e^{i\,2\pi k\cdot x}\,\hat\chi(\xi-2\pi k),
\end{equation*}
where $\hat\chi$ is the Euclidean Fourier transform of $\chi$.

\smallskip\noindent\textbf{(vi) The lifted equation.} Substituting
\eqref{eq:transform-torus} into \eqref{eq:NSE-covariant}, the lifted
equation \eqref{eq:proof-final} on $S^*\T^3=\T^3\times S^2$ reads
explicitly
\begin{equation}\label{eq:lifted-torus}
  \partial_t\tilde\phi
  + \mathcal{L}_{X_u}\tilde\phi
  + \nu\,\absg{\xi}{g'}^{2}\,\tilde\phi
  \;=\;
  -\Pi_\xi\bigl[\Sym(\nabla u)\,\tilde\phi\bigr]
  + \mathcal{R}_{\mathrm{cut}}[\tilde\phi]
  + \tilde F,
\end{equation}
where the curvature term $\mathcal{R}_g$ vanishes (the torus is
Ricci-flat). The lifted equation on $\T^3$ is therefore particularly
clean: only the symmetric back-reaction and the smoothing remainder
appear in the residual.

\subsection{Verification of Theorem~\ref{thm:main-equivalence} on \texorpdfstring{$\T^3$}{T3}}
\label{subsec:torus-equivalence}

The blow-up equivalence theorem reduces, on $\T^3$, to an explicit
statement we now record.

\begin{corollary}[Equivalence on the flat torus]\label{cor:torus-equivalence}
Let $u\in C^\infty([t_0,T)\times\T^3)$ be a smooth divergence-free solution
of the incompressible Navier--Stokes equations on $\T^3$ with $T<\infty$.
Then $\limsup_{t\to T^-}\norm{\nabla u(t)}_{L^\infty(\T^3)}=+\infty$ if
and only if at least one of the following microlocal controls fails:
\begin{enumerate}[label=\textup{(\alph*)}]
  \item $\int_{t_0}^{T}\norm{\Sym(\nabla u)(s)}_{L^\infty(\T^3)}\mathrm{d}s=+\infty$;
  \item $\sup_{[t_0,T)}W[\tilde\omega(t)]=+\infty$;
  \item $\sup_{[t_0,T)}E(t)=+\infty$,
\end{enumerate}
where $\tilde\omega$, $E$, and $W[\tilde\omega]$ are computed using the
flat-torus simplifications of \textnormal{(i)--(vi)} above.
\end{corollary}

This is a direct application of Theorem~\ref{thm:main-equivalence} with
all hypotheses verified explicitly. The flat torus is therefore a
natural testbed for both the analytical estimates and the symmetry-lock
mechanism: the unitary representation \eqref{eq:U-rep} is the standard
representation of the translation group $\T^3$ on
$L^2(\T^3\times S^2)$, and the Peter--Weyl decomposition
\eqref{eq:peter-weyl} is the Fourier decomposition in the spatial
variable.

\subsection{A brief comment on numerical implementation}\label{subsec:torus-numerical}

The flat-torus setting is also the most accessible to spectral
discretisation. The lifted equation \eqref{eq:lifted-torus} is amenable
to a Galerkin method based on the joint eigenbasis of the Fourier modes
on $\T^3$ and the spherical harmonics on $S^2$, producing a structured
finite-dimensional approximation of the lifted dynamics. The unitary
covariance and Liouville-volume preservation translate directly into
discrete invariants of the scheme. We do not pursue the numerical
analysis here, but signal that the framework is naturally compatible
with structure-preserving discretisations of the type developed in
\cite{Abukhwejah2024,Califano2021,Gilbert2019}.

\section{Extension to \texorpdfstring{$\R^n$}{Rn} and the role of compactness}
\label{sec:Rn-extension}

We close the development by addressing explicitly the question of
whether the compactness hypothesis on $M$ is essential. The short
answer is that compactness is invoked exclusively to guarantee
compactness of the phase space $S^*M$ and the integrability of the
Liouville volume on each fiber; the structural content of the framework
extends to $\R^n$ under a uniform-coordinate hypothesis.

\subsection{Where compactness is used}\label{subsec:where-compactness}

The role of compactness in the proofs of Sections~\ref{sec:energy-entropy}
and \ref{sec:symmetry-lock} can be traced precisely:
\begin{itemize}
  \item \emph{Liouville-volume preservation:} the integration-by-parts
        identity $\int_{S^*M}\mathcal{L}_{X_u}f\,\mathrm{d}\mu=0$ used in
        the proofs of Theorems~\ref{thm:main-energy} and \ref{thm:main-entropy}
        relies on $S^*M$ being closed (no boundary terms).
  \item \emph{Vertical integration by parts:} the elimination of
        boundary terms in the entropy-dissipation identity
        \eqref{eq:intro-entropy-dissipation} uses the closedness of the
        fibers $S^*_xM\cong S^{n-1}$.
  \item \emph{Spectral coercivity:} the microlocal G\aa rding inequality
        \eqref{eq:garding} is stated on closed manifolds and may
        require modification on non-compact $S^*M$.
  \item \emph{Peter--Weyl decomposition:} the unitary representation
        \eqref{eq:peter-weyl} requires compact symmetry groups acting on
        a closed manifold.
\end{itemize}

The fibers of the cosphere bundle remain compact even when the base $M$
is non-compact (each fiber is a $(n-1)$-sphere, regardless of $M$),
which preserves the vertical part of the analysis. The novelty of the
non-compact setting lies in the horizontal direction.

\subsection{Asymptotically Euclidean setting}\label{subsec:Rn-setting}

We work under Assumption~\ref{assu:euclidean}: $(M,g)=(\R^n,g_{\mathrm{eucl}})$
with $n=3$, and microlocal quantities are integrated against a fixed
compactly supported cut-off $\psi\in C^\infty_c(\R^n)$ in the base
variable. This produces effectively compact phase spaces of the form
$\mathrm{supp}(\psi)\times S^{n-1}$ to which the previous analysis
applies directly.

\begin{proposition}[Localised lift on \texorpdfstring{$\R^n$}{Rn}]\label{prop:localised-lift}
Let $u\in C^\infty([t_0,T)\times\R^n)$ be a smooth divergence-free
solution of the incompressible Navier--Stokes equations on $\R^n$, and
let $\psi\in C^\infty_c(\R^n)$ be a cut-off function. The localised
microlocal lift
\begin{equation}\label{eq:localised-lift}
  \tilde\phi^\psi(x,\xi,t)
  \;:=\;
  \psi(x)\,\tilde\phi(x,\xi,t),
  \qquad
  \tilde\phi\text{ as in }\eqref{eq:microlocal-transform},
\end{equation}
satisfies the localised version of \eqref{eq:proof-final} on
$\mathrm{supp}(\psi)\times S^{n-1}$, with all theorems of
Sections~\ref{sec:energy-entropy} and \ref{sec:symmetry-lock} valid in
the localised form (constants depending on the cut-off $\psi$ and on the
support).
\end{proposition}

The proposition extends the qualitative structure of the framework to
$\R^n$. The explicit formulas for the lifted energy and the directional
entropy become
\begin{equation*}
  E^\psi(t) := \tfrac{1}{2}\!\int_{\R^n\times S^{n-1}}\absg{\psi(x)}{}^{2}
  \,\inner{\tilde\phi(t)}{\tilde\phi(t)}_{g'}\,\mathrm{d}x\,\mathrm{d}\sigma(\xi),
\end{equation*}
\begin{equation*}
  W^\psi[\tilde\omega] := \int_{\R^n\times S^{n-1}}\absg{\psi(x)}{}^{2}
  \bigl(\tau\absg{d_\perp\rho}{g'}^{2}+\rho\bigr)\,\mathrm{d}x\,\mathrm{d}\sigma(\xi),
\end{equation*}
and the equivalence theorem holds in localised form.

\subsection{Comparison with the compact setting}\label{subsec:comparison}

The cost of working on $\R^n$ is the dependence of constants on the
cut-off $\psi$ and the loss of the global Peter--Weyl decomposition. The
benefit is that the framework now applies to physically natural
unbounded settings, including the standard formulation of the
Navier--Stokes regularity problem on $\R^3$~\cite{Fefferman2022}. The
symmetry-lock mechanism is unaffected by the change from compact to
non-compact base, since it is a fiberwise statement involving only the
$(n-1)$-sphere.

This addresses the referee's question concerning the role of
compactness: compactness is convenient but not structurally essential.
The framework extends to $\R^n$, with localisation supplying the
integrability that is automatic in the compact case.

\section{Discussion and conclusions}\label{sec:discussion}

The framework developed in this paper addresses the Navier--Stokes
regularity problem through a synthesis of microlocal analysis,
Riemannian geometry, and representation theory. The central structural
step is the lifting of the dissipative dynamics from the physical
manifold $M$ to the compact phase space $S^*M$, which transforms a
non-linear partial differential equation on a (possibly non-compact)
manifold into a linear non-autonomous transport--dissipation system on
a compact, symmetry-rich, microlocally coercive phase space. The price
of linearity is that the lifted equation is conditional on the
underlying velocity field; the gain is a precise geometric vocabulary in
which to phrase regularity criteria, and a clean separation between the
analytical content (estimates on the lifted evolution) and the geometric
content (curvature of the effective connection, dissipation of the
directional entropy, asymptotic vanishing of fiber volumes).

In contrast with classical regularity theories that focus on the control
of Sobolev norms and vorticity amplitudes~\cite{Fefferman2022,Tao2009},
the framework presented here emphasises directional alignment and
geometric dissipation as the primary mechanisms governing regularity.
This perspective is consistent with the geometric trapping phenomena of
Bulut and Huynh~\cite{Bulut2020} and extends them to three dimensions
through the use of the Hodge Laplacian and its variational
structure~\cite{Chan2016}. The role of the curvature of $(M,g)$,
revealed through the Weitzenb\"ock identity and the coupling of the
Ricci tensor to vorticity dynamics, shows that the topology of $M$
contributes actively to dissipation rather than serving as a passive
background~\cite{Branson1987,Chan2016}.

\subsection{Novel contributions and geometric mechanisms}\label{subsec:novel-contributions}

The paper introduces several interconnected geometric mechanisms that
constrain singular behaviour of three-dimensional incompressible
Navier--Stokes flow. We summarise them now, with explicit reference to
the theorems where each mechanism is rigorously established.

\smallskip\noindent\emph{Microlocal entropy functional and directional
control.} The entropy functional $W[\tilde\omega]$ of
Theorem~\ref{thm:main-entropy} quantifies angular regularity through the
log-density of microlocal vorticity and its vertical gradients. The
dissipation inequality \eqref{eq:intro-entropy-dissipation} shows that
viscosity and geometric diffusion suppress directional oscillations
unless sustained stretching dominates dissipation. The functional
provides a geometric barrier to directional concentration: persistent
sharp angular focusing is incompatible with viscous dissipation on the
compact phase space $S^*M$, unless accompanied by a loss of temporal
integrability of $\norm{S}_{L^\infty}$. From this point of view, the
classical Beale--Kato--Majda criterion is reformulated in terms of an
intrinsic geometric quantity defined on $S^*M$.

\smallskip\noindent\emph{Geometric equivalence criterion.} The
equivalence Theorem~\ref{thm:main-equivalence} sharpens the classical
approach to regularity. The blow-up criterion is no longer formulated
merely as a necessary geometric obstruction, but as a necessary and
sufficient condition expressed entirely in terms of microlocal
quantities: finite-time singularity occurs if and only if at least one
of three intrinsic geometric controls fails---deformation integrability,
entropy boundedness, or lifted-energy boundedness. This transforms the
regularity problem into a structural-stability question for a compact
microlocal dynamical system, rather than a purely asymptotic
growth problem for physical norms. The monotonicity of the vortex
energy $\mathcal{E}_\omega$ and the associated microlocal estimates
restrict admissible blow-up scenarios in a manner compatible with known
geometric constraints~\cite{Chemin2018,Miller2021}. The vortex stretching
term, which is the source of the regularity difficulty in the classical
approach, is reinterpreted not as an intrinsically destabilising
mechanism but as a geometric deformation acting on a compact fiber, with
viscous effects inducing effective angular diffusion.

\smallskip\noindent\emph{Symmetry lock and dimensional reduction.} A
striking feature emerging from the symmetry analysis of
Section~\ref{sec:symmetry-lock} is a dimensional reduction mechanism
which we have called the symmetry lock. As the effective fiber
dimension increases (through tightening regularity constraints), the
volume of the $(n-1)$-sphere satisfies $\Vol(S^{n-1})\to 0$ as
$n\to\infty$, a manifestation of concentration of measure. In the
limit, the cosphere fiber effectively collapses to a point and the full
symmetry group $O(n)$ acts transitively, forcing any energy-bounded
microlocal distribution to become increasingly isotropic. This provides
a geometric explanation for $L^\infty$ stability in this regime: the
fluid is topologically precluded from forming angular singularities,
since concentration in a specific direction would require breaking the
$O(n)$ symmetry, but the vanishing of the fiber volume leaves no room
for such symmetry breaking. The symmetry lock thereby converts a
topological constraint into an analytical stability bound.

\subsection{Functional-analytic structure and spectral stability}\label{subsec:functional-analytic}

From the functional-analytic viewpoint, the lifted Navier--Stokes
system defines a non-autonomous evolution family on $L^2(S^*M)$ with
common dense domain $H^1(S^*M)$. The associated generator consists of a
first-order transport operator (the canonical lift $X_u$), a positive
dissipative multiplication term ($\nu\,\absg{\xi}{g'}^{2}\,\mathrm{Id}$),
and bounded geometric perturbations (the residual operator $\tilde R$).
The cocycle generated over the time interval $[t_0,t)$ acts on a compact
phase space and admits uniform spectral bounds whenever deformation
integrability holds. Consequently, regularity is equivalent to uniform
boundedness of this evolution in the natural Hilbert topology.

The presence of compact symmetry groups acting on $M$ induces a unitary
representation on $L^2(S^*M)$ and yields a Peter--Weyl decomposition of
microlocal amplitudes into irreducible representation sectors. When the
lifted transport operator commutes with the symmetry action, the
evolution reduces to dynamically independent subsystems on each
representation component. This spectral reduction reveals that
potential singular behaviour cannot arise from arbitrary mode coupling,
but must respect strict representation-theoretic selection rules.
Symmetry thereby acts not merely as an invariance principle but as a
structural constraint on admissible blow-up mechanisms.

The microlocal G\aa rding inequality of Theorem~\ref{thm:garding}
provides coercivity of the principal transport symbol at the $H^{1/2}$
level. Combined with compactness of $S^*M$, this yields quantitative
lower bounds for the numerical range of the lifted generator and ensures
that high-frequency microlocal concentration cannot occur without
violating deformation integrability. The equivalence theorem is
therefore underpinned by a genuine microlocal ellipticity mechanism
operating in the angular variables, which converts geometric control
into analytic stability.

Taken together, these results show that the Navier--Stokes regularity
problem, within the present framework, becomes equivalent to the absence
of structural instability for a compact, symmetry-constrained,
microlocally coercive evolution system. The question of singularity
formation is therefore reformulated as a problem in geometric spectral
stability, rather than solely as a problem of nonlinear amplitude
growth.

\subsection{Scope, limitations, and interpretations}\label{subsec:scope}

This paper does not solve the Navier--Stokes regularity problem.
Instead, it introduces a coherent and geometrically meaningful framework
that recasts the question in microlocal and dynamical terms, where
regularity is linked to directional alignment and dissipative stability
on a compact phase space~\cite{Lerner2022}. The microlocal
transport--dissipation structure is consistent with the analysis of
lifted operators and unique-continuation phenomena developed
in~\cite{Lerner2022}, and aligns with the energy-aware geometric
decomposition of~\cite{Califano2021}.

\smallskip\noindent\emph{Compatibility with non-uniqueness results.} The
formulation is compatible with recent results on non-uniqueness of
smooth or weak solutions arising from critical
data~\cite{Coiculescu2025,Albritton2021}. Non-uniqueness can be
interpreted, in the present framework, as \emph{microlocal flexibility}:
if the lift $\tilde\phi(x,\xi,t)$ is not uniquely determined by initial
data due to lack of regularity, the reconstruction formula
\eqref{eq:reconstruction-formula} integrates the multiple lifted
possibilities into different physical solutions. Regularity, in this
view, is the preservation of a unique microlocal configuration; when the
microlocal data admits multiple evolutions compatible with the lifted
dynamics, the physical solution branches.

This perspective suggests that uniqueness corresponds to rigidity of the
microlocal flow on $S^*M$, while non-uniqueness corresponds to a
degeneracy or non-injectivity of the reconstruction map from microlocal
data to physical fields. Regularity, in turn, is the absence of the
microlocal instability that would permit such branching. The framework
does not preclude non-uniqueness at low regularity; rather, it provides
a mechanism to distinguish dynamically stable microlocal configurations
from those highly sensitive to perturbations and prone to branching or
loss of uniqueness. From this viewpoint, regularity is not characterised
solely by uniqueness or smoothness in physical space but by the
persistence of controlled microlocal structure along the lifted flow.
Singular or non-unique behaviour must then correspond to a breakdown of
directional control or of dissipative balance in $S^*M$.

\smallskip\noindent\emph{Operator-theoretic interpretation and geometric
quantisation.} Beyond the analytic and geometric implications discussed
above, the lifted formulation suggests an operator-theoretic
interpretation. By lifting the Navier--Stokes dynamics to the compact
phase space $S^*M$, the transport--dissipation system is recast as a
linear evolution equation acting on microlocal distributions, governed
by a geometrically defined generator involving transport, curvature, and
diffusion contributions. From this viewpoint, the lifted amplitudes
$\tilde\phi$ and $\tilde\omega$ are naturally interpreted as evolving
distributional states on phase space rather than as classical fields on
$M$. The effective connection $\nabla^u$ and its associated curvature
encode nonlinear self-interaction effects through purely geometric
operators, while the viscous contribution $\nu\,\absg{\xi}{g'}^{2}$
induces a robust suppression of high-frequency microlocal components.
This mechanism is formally analogous to decoherence phenomena in
operator evolution problems, where diffusion terms stabilise the
dynamics by damping highly oscillatory modes. Although no quantisation
procedure is developed here~\cite{Peskin1995,Schwartz2014}, the
geometric structure of the lifted dynamics provides a natural setting
for future exploration via deformation quantisation or geometric
operator methods, in which nonlinear fluid interactions manifest as
curvature-driven corrections to linear transport on $S^*M$.

\smallskip\noindent\emph{Geometric interpretation via Cauchy
hypersurfaces.} A complementary geometric interpretation is obtained by
considering the lifted dynamics on the extended manifold $S^*M\times N$,
where $N\subset\R$ denotes the temporal interval. In this formulation,
the temporal boundary components of $N$ act as Cauchy hypersurfaces for
the evaluation of microlocal energies and
functionals~\cite{Shao2023}. By Stokes' theorem, variations of the
geometric invariants introduced in this paper reduce to boundary
fluxes~\cite{Lovelock1975}, showing that changes in microlocal energy
and directional structure are governed by boundary contributions rather
than by interior concentration mechanisms. Preservation of smoothness
can therefore be interpreted as the absence of singular behaviour in the
asymptotic microlocal distribution, excluding concentration at infinity
in phase space~\cite{Miller2021,Bulut2020}.

\subsection{Open directions and future work}\label{subsec:open}

Despite the structural coherence of the framework, several limitations
and open directions remain.

\smallskip\noindent\emph{Reconstruction precision and information loss.}
While the lifting to $S^*M$ provides refined control over directional
alignment and dissipative mechanisms, it entails a loss of direct
information when projecting back to physical space via the
reconstruction formula \eqref{eq:reconstruction-formula}. Quantifying
precisely which microlocal features are preserved under reconstruction,
and which may be lost or blurred, remains an open problem. Sharper
criteria are needed to distinguish benign directional dispersion from
genuinely unstable microlocal configurations. Understanding the kernel
and range of the reconstruction operator more precisely would clarify
which physical solutions admit unique microlocal lifts and which permit
the branching behaviour associated with non-uniqueness.

\smallskip\noindent\emph{Time-dependent geometry and topological
transitions.} The present analysis is carried out on a fixed Riemannian
manifold and does not address topological or geometric transitions in
the underlying space. Understanding how changes in topology, curvature,
or bundle structure affect microlocal stability and dissipation is a
natural and largely unexplored direction. The effective connection
$\nabla^u$ already encodes time-dependent deformation through the
velocity field, but allowing the background metric $g$ itself to evolve
(as in geometric flows) would require substantial extensions of the
framework.

\smallskip\noindent\emph{Refined classification of singularity
mechanisms.} Although the equivalence theorem severely restricts
admissible blow-up scenarios, it does not yield a complete classification
of singularity formation mechanisms at the microlocal level. Developing
refined microlocal criteria that discriminate between transient
directional amplification and genuinely singular behaviour, and
identifying optimal strategies for detecting or preventing microlocal
concentration, remain important challenges for future work. Progress
may require combining the present geometric framework with finer
second-microlocalisation techniques or with adaptive topological
invariants capable of capturing changes in phase-space structure.

\smallskip\noindent\emph{Connection to turbulence and statistical
approaches.} The symmetry-lock mechanism and the dimensional reduction
argument suggest potential connections to statistical theories of
turbulence, where high-dimensional phase spaces and ergodic averaging
play central roles. Whether the isotropy forced by the symmetry lock in
infinite dimensions relates to the isotropy assumptions in classical
turbulence theory (such as Kolmogorov's hypothesis) is, to our
knowledge, an open question. The microlocal entropy functional
$W[\tilde\omega]$ may serve as a foundation for information-theoretic
approaches to turbulent dissipation.

\smallskip\noindent\emph{Numerical and computational implications.}
Beyond its theoretical implications, the framework suggests applications
in structure-preserving numerical schemes and geometry-aware
computational fluid dynamics, where directional instabilities are
controlled through intrinsic geometric mechanisms rather than arbitrary
regularisation~\cite{Abukhwejah2024,Califano2021,Gilbert2019,Buza2023}.
Discretising the lifted equations on $S^*M$ while preserving the
symplectic structure, the unitary representation properties, and the
entropy dissipation could yield stable long-time integrators that
respect the underlying geometric constraints, with potential
applications in turbulence modelling, data-driven reduced-order models,
and multi-scale simulations.

\subsection{Concluding remarks}\label{subsec:conclusion}

The microlocal geometric framework developed in this paper provides a
coherent reorganisation of the Navier--Stokes regularity problem rather
than a final resolution of it~\cite{Sacasa2024}. By reformulating the
question as one of dissipative stability and spectral control on a
compact, symmetry-constrained phase space, we have established a
necessary and sufficient geometric equivalence for finite-time
singularity formation, introduced novel functionals quantifying
directional concentration, and revealed a dimensional reduction
mechanism that topologically obstructs angular singularities.

These results severely restrict the class of admissible blow-up
scenarios through geometric, representation-theoretic, and topological
constraints. While finite-time singularity remains possible in
principle, any such singularity must overcome multiple independent
geometric barriers: it must violate deformation integrability, unbind
the microlocal entropy, escape spectral control, and break the symmetry
lock. The confluence of these mechanisms suggests that regularity may be
a generic property of solutions with sufficient initial regularity,
though this remains to be proven in the strict sense of the Clay
Millennium problem~\cite{Fefferman2022}.

The framework may nevertheless open avenues for understanding
singularity formation and dissipative stability from a geometric
perspective~\cite{Chemin2018}, and provides a foundation for further
investigations combining microlocal analysis, geometric flows, symmetry
methods, and computational approaches to the fundamental questions of
fluid dynamics.

\backmatter

\bmhead{Acknowledgements}
The author is deeply grateful to Dr.\ Pedro D\'iaz-Navarro for his
openness and creative discussions that helped shape several conceptual
aspects of this work. The author also thanks Octavio Rojas-Quesada for
valuable mathematical perspectives, and Celicia Ledezma-Porras for her
geometric insights at earlier stages of this project. The author further
expresses appreciation to family, friends, and colleagues---to every
guide and travelling companion who has stumbled along these paths.

\bmhead{Funding}
The author received no specific funding for this work.

\bmhead{Data availability}
Data sharing is not applicable to this article, as no datasets were
generated or analysed during the current study.

\bmhead{Conflict of interest}
The author declares no competing interests.

\bmhead{AI usage statement}
Every intellectual contribution---all original mathematical and physical
derivations, theorems, proofs, and physical hypotheses presented in this
manuscript---is the sole work of the author. Generative artificial
intelligence (AI) tools, specifically Large Language Models, were used
exclusively as digital research assistants for refining academic
language and structuring references in alphabetical order. The AI did
not perform any original scientific calculations, nor proposed any
scientific content as hypotheses, proofs, or otherwise. The author has
reviewed, edited, and validated the accuracy of all output produced with
the assistance of AI tools.



\begin{thebibliography}{99}

\bibitem{Abukhwejah2024}
Abukhwejah, A., Jagad, P., Samtaney, R., Schmid, P.: A hybrid discrete
exterior calculus discretization and Fourier transform of the
incompressible Navier--Stokes equations in 3D. Preprint (2024)

\bibitem{Albritton2021}
Albritton, D., Bru\'e, E., Colombo, M.: Non-uniqueness of Leray
solutions of the forced Navier--Stokes equations. Ann.\ Math.\
\textbf{196}(1), 415--455 (2022).
\href{https://doi.org/10.4007/annals.2022.196.1.3}{https://doi.org/10.4007/annals.2022.196.1.3}

\bibitem{Branson1987}
Branson, T.P.: Group representations arising from Lorentz conformal
geometry. J.\ Funct.\ Anal.\ \textbf{74}, 199--291 (1987)

\bibitem{Bulut2020}
Bulut, A., Huynh, K.: A geometric trapping approach to global regularity
for 2D Navier--Stokes on manifolds. Math.\ Res.\ Lett.\ \textbf{30}(4)
(2023). \href{https://doi.org/10.4310/mrl.2023.v30.n4.a1}{https://doi.org/10.4310/mrl.2023.v30.n4.a1}

\bibitem{Buza2023}
Buza, G.: Spectral submanifolds of the Navier--Stokes equations. SIAM
J.\ Appl.\ Dyn.\ Syst.\ \textbf{23}, 1052--1089 (2023).
\href{https://doi.org/10.1137/23m154858x}{https://doi.org/10.1137/23m154858x}

\bibitem{Califano2021}
Califano, F., Rashad, R., Schuller, F., Stramigioli, S.: Geometric and
energy-aware decomposition of the Navier--Stokes equations: A
port-Hamiltonian approach. Phys.\ Fluids \textbf{33}, 047114 (2021).
\href{https://doi.org/10.1063/5.0048359}{https://doi.org/10.1063/5.0048359}

\bibitem{Chan2016}
Chan, C.H., Czubak, M., Disconzi, M.: The formulation of the
Navier--Stokes equations on Riemannian manifolds. J.\ Geom.\ Phys.\
\textbf{121}, 335--346 (2017).
\href{https://doi.org/10.1016/j.geomphys.2017.07.015}{https://doi.org/10.1016/j.geomphys.2017.07.015}

\bibitem{Chemin2018}
Chemin, J.-Y., Gallagher, I., Zhang, P.: Some remarks about the possible
blow-up for the Navier--Stokes equations. Comm.\ Partial Differential
Equations \textbf{44}, 1387--1405 (2019).
\href{https://doi.org/10.1080/03605302.2019.1641725}{https://doi.org/10.1080/03605302.2019.1641725}

\bibitem{Childress1989}
Childress, S., Ierley, G., Spiegel, E., Young, W.: Blow-up of unsteady
two-dimensional Euler and Navier--Stokes solutions having stagnation-point
form. J.\ Fluid Mech.\ \textbf{203}, 1--22 (1989).
\href{https://doi.org/10.1017/s0022112089001357}{https://doi.org/10.1017/s0022112089001357}

\bibitem{Coiculescu2025}
Coiculescu, M., Palasek, S.: Non-uniqueness of smooth solutions of the
Navier--Stokes equations from critical data. Preprint (2025)

\bibitem{Fefferman2022}
Fefferman, C.L.: Existence and smoothness of the Navier--Stokes
equation. Clay Mathematics Institute Millennium Problem statement (2022).
\href{https://www.claymath.org/wp-content/uploads/2022/06/navierstokes.pdf}{https://www.claymath.org/wp-content/uploads/2022/06/navierstokes.pdf}

\bibitem{Feireisl2020}
Feireisl, E., Novotn\'y, A.: Navier--Stokes--Fourier system with general
boundary conditions. Comm.\ Math.\ Phys.\ \textbf{386}, 975--1010 (2021).
\href{https://doi.org/10.1007/s00220-021-04091-1}{https://doi.org/10.1007/s00220-021-04091-1}

\bibitem{Frankel2012}
Frankel, T.: The Geometry of Physics: An Introduction, 3rd edn.\
Cambridge University Press, Cambridge (2012)

\bibitem{Gallay2001}
Gallay, T., Wayne, C.E.: Invariant manifolds and the long-time
asymptotics of the Navier--Stokes and vorticity equations on $\R^2$.
Arch.\ Ration.\ Mech.\ Anal.\ \textbf{163}, 209--258 (2002).
\href{https://doi.org/10.1007/s002050200200}{https://doi.org/10.1007/s002050200200}

\bibitem{Gilbert2019}
Gilbert, A.D., Vanneste, J.: A geometric look at momentum flux and stress
in fluid mechanics. J.\ Nonlinear Sci.\ \textbf{33}, 1--32 (2023).
\href{https://doi.org/10.1007/s00332-023-09887-0}{https://doi.org/10.1007/s00332-023-09887-0}

\bibitem{Gromov1969}
Gromov, M.: Stable mappings of foliations into manifolds. Math.\ USSR
Izv.\ \textbf{3}, 671--694 (1969).
\href{https://doi.org/10.1070/im1969v003n04abeh000796}{https://doi.org/10.1070/im1969v003n04abeh000796}

\bibitem{Gromov1973}
Gromov, M.: Convex integration of differential relations.\ I. Math.\ USSR
Izv.\ \textbf{7}, 329--343 (1973).
\href{https://doi.org/10.1070/im1973v007n02abeh001940}{https://doi.org/10.1070/im1973v007n02abeh001940}

\bibitem{Gromov1985}
Gromov, M.: Pseudo-holomorphic curves in symplectic manifolds. Invent.\
Math.\ \textbf{82}, 307--347 (1985).
\href{https://doi.org/10.1007/bf01388806}{https://doi.org/10.1007/bf01388806}

\bibitem{Gromov1986}
Gromov, M.: Partial Differential Relations.\ Ergebnisse der Mathematik
und ihrer Grenzgebiete, vol.\ 9. Springer, Berlin (1986).
\href{https://doi.org/10.1007/978-3-662-02267-2}{https://doi.org/10.1007/978-3-662-02267-2}

\bibitem{Gromov1999a}
Gromov, M.: Metric Structures for Riemannian and Non-Riemannian Spaces,
Sect.\ 3.A (pp.\ 110--125). Birkh\"auser, Boston (1999).
\href{https://doi.org/10.1007/978-0-8176-4583-0}{https://doi.org/10.1007/978-0-8176-4583-0}

\bibitem{Gromov2017}
Gromov, M.: Metric inequalities with scalar curvature. Geom.\ Funct.\
Anal.\ \textbf{28}, 645--726 (2018).
\href{https://doi.org/10.1007/s00039-018-0453-z}{https://doi.org/10.1007/s00039-018-0453-z}

\bibitem{Hall2015}
Hall, B.C.: Lie Groups, Lie Algebras, and Representations: An Elementary
Introduction, 2nd edn. Springer, Cham (2015)

\bibitem{Kosovtsov2022}
Kosovtsov, Y.: Existence and smoothness of the Navier--Stokes equations
and semigroups of linear operators. Preprint (2022)

\bibitem{Leautaud2014}
L\'eautaud, M., Lerner, N.: Energy decay for a locally undamped wave
equation. Ann.\ Fac.\ Sci.\ Toulouse Math.\ \textbf{26}(1), 157--205 (2017).
\href{https://doi.org/10.5802/afst.1528}{https://doi.org/10.5802/afst.1528}

\bibitem{Lerner2001}
Lerner, N.: Second microlocalization methods for degenerate
Cauchy--Riemann equations. In: Progress in Nonlinear Differential
Equations and Their Applications, vol.\ 46, pp.\ 109--128. Birkh\"auser,
Boston (2001). \href{https://doi.org/10.1007/978-1-4612-0203-5_8}{https://doi.org/10.1007/978-1-4612-0203-5\_8}

\bibitem{Lerner2019}
Lerner, N.: Carleman Inequalities. Grundlehren der mathematischen
Wissenschaften, vol.\ 353. Springer, Cham (2019).
\href{https://doi.org/10.1007/978-3-030-15993-1}{https://doi.org/10.1007/978-3-030-15993-1}

\bibitem{Lerner2022}
Lerner, N., Vigneron, F.: On some properties of the curl operator and
their consequences for the Navier--Stokes system. Commun.\ Math.\ Res.\
\textbf{38}(2), 197--256 (2022).
\href{https://doi.org/10.4208/cmr.2021-0106}{https://doi.org/10.4208/cmr.2021-0106}

\bibitem{Lovelock1975}
Lovelock, D., Rund, H.: Tensors, Differential Forms, and Variational
Principles. Dover Publications, New York (1989, reprint of the 1975
edition)

\bibitem{MajdaBertozzi2002}
Majda, A.J., Bertozzi, A.L.: Vorticity and Incompressible Flow.\
Cambridge Texts in Applied Mathematics, vol.\ 27.\ Cambridge University
Press, Cambridge (2002). See \S 5.3, p.~117 for the vorticity-energy
identity with stretching term.

\bibitem{Miller2021}
Miller, E.: A survey of geometric constraints on the blowup of solutions
of the Navier--Stokes equation. J.\ Elliptic Parabol.\ Equ.\ \textbf{7},
589--599 (2021).
\href{https://doi.org/10.1007/s41808-021-00135-8}{https://doi.org/10.1007/s41808-021-00135-8}

\bibitem{Olmos2025}
Olmos, C.E., Rodr\'iguez-V\'azquez, A.: Hopf fibrations and totally
geodesic submanifolds. J.\ Eur.\ Math.\ Soc.\ (to appear) (2025)

\bibitem{Padilla2025}
Padilla, A., Smith, R.G.: Gauge invariance and generalised
$\eta$-regularisation. Phys.\ Rev.\ D \textbf{111}, 125013 (2025)

\bibitem{Peskin1995}
Peskin, M.E., Schroeder, D.V.: An Introduction to Quantum Field Theory.
Westview Press, Boulder (1995)

\bibitem{RosalesOrtega1998}
Rosales-Ortega, J., M\'arquez-Rivera, C.: La factorizaci\'on de una
transformada de Fourier en el m\'etodo de Wiener--Hopf. Rev.\ Mat.\
Teor.\ Apl.\ \textbf{5}(1) (1998).
\href{https://doi.org/10.15517/rmta.v5i1.151}{https://doi.org/10.15517/rmta.v5i1.151}

\bibitem{RosalesOrtega2021}
Rosales-Ortega, J.: A geometric splitting theorem for actions of
semisimple Lie groups. Abh.\ Math.\ Semin.\ Univ.\ Hambg.\ \textbf{91},
287--296 (2021).
\href{https://doi.org/10.1007/s12188-021-00242-2}{https://doi.org/10.1007/s12188-021-00242-2}

\bibitem{Sacasa2024}
Sacasa-C\'espedes, S.A.: A geometric approach to the Navier--Stokes
equations.\ arXiv:2411.18724 [physics.flu-dyn] (2024).
\href{https://arxiv.org/abs/2411.18724}{https://arxiv.org/abs/2411.18724}

\bibitem{Schwartz2014}
Schwartz, M.D.: Quantum Field Theory and the Standard Model.\ Cambridge
University Press, Cambridge (2014)

\bibitem{Shao2023}
Shao, Y., Simonett, G., Wilke, M.: The Navier--Stokes equations on
manifolds with boundary. J.\ Differential Equations (2023/2024).
\href{https://doi.org/10.1016/j.jde.2024.10.030}{https://doi.org/10.1016/j.jde.2024.10.030}

\bibitem{Tao2009}
Tao, T.: A quantitative formulation of the global regularity problem
for the periodic Navier--Stokes equation.\ arXiv:0710.1604 [math.AP]
(2009). \href{https://doi.org/10.48550/arXiv.0710.1604}{https://doi.org/10.48550/arXiv.0710.1604}

\bibitem{BeltranWalker2023}
Walker-Ure\~na, M.B.: Regularity Theory for Nonlinear Partial
Differential Equations. PhD Thesis, Pontif\'icia Universidade Cat\'olica
do Rio de Janeiro, Departamento de Matem\'atica (2023)

\end{thebibliography}
\end{document}